\documentclass{aa}  
\usepackage{graphicx}
\usepackage{graphics}
\usepackage[varg]{txfonts}
\usepackage[colorlinks=true,allcolors=blue]{hyperref} 
\usepackage{enumerate}
\usepackage{amsmath}
\usepackage{natbib}
\usepackage{comment}
\usepackage[para]{threeparttable}
\usepackage{xcolor}
\usepackage{siunitx}

\begin{document} 

   \title{Relating spatially resolved optical attenuation, dust and gas in nearby galaxies}

   \subtitle{}

   \author{
            E. D. Paspaliaris\inst{1}\and
            S. Bianchi\inst{1}\and
            E. Corbelli\inst{1}\and
            A. Concas\inst{2,3}
          }

   \institute{INAF - Osservatorio Astrofisico di Arcetri, Largo E. Fermi 5, 50125, Florence, Italy\\
   email: \href{mailto:evangelos.paspaliaris@inaf.it}{evangelos.paspaliaris@inaf.it}, \href{mailto:edpaspaliaris@gmail.com}{edpaspaliaris@gmail.com}
   \and
   European Southern Observatory, Karl-Schwarzschild-Strasse 2, 85748 Garching bei München, Germany
   \and
   Scuola Normale Superiore, Piazza dei Cavalieri 7, 50126 Pisa, Italy}
   
   \date{Received Apr 14, 2025 / Accepted Aug 15, 2025}

  \abstract
   {}
   {The purpose of the present study is to relate the optical attenuation  inferred by the Balmer decrement, $A_{V,\mathrm{BD}}$, and by the spectral energy distribution (SED)-fitting, $A_{V,\mathrm{SED}}$, to the dust distribution and gas surface density throughout the disc of galaxies, down to scales smaller than 0.5 kpc.}
   {We investigate five nearby \textit{Herschel}-detected star-forming spiral galaxies with available far-ultraviolet to sub-millimetre observations, along with atomic and molecular gas surface density maps and optical integral-field spectroscopic data. We use the \texttt{CIGALE} SED-fitting code to map the dust mass surface density ($\Sigma _\mathrm{dust}$) and  $A_{V,\mathrm{SED}}$ of different stellar populations. For each pixel, we independently estimate the attenuation from the Balmer decrement.} 
   {We find that both $\Sigma _\mathrm{dust}$ and $A_{V,\mathrm{BD}}$ trace better the molecular and total gas mass surface density, rather than the atomic gas. Since regions sampled in this study have high molecular fractions, atomic gas surface densities, indicative of molecular gas shielding  layers, decrease as the mean dust-to-gas ratio increases from galaxy to galaxy. The fitted attenuation towards the young stellar population, $A^\mathrm{young}_{V,\mathrm{SED}}$, is in good agreement with $A_{V,\mathrm{BD}}$ and it can then be used to trace the attenuation in star forming galaxies where integral-field observations are not available. We estimate the ratio of $A_{V,\mathrm{BD}}$ over the total stellar $A_{V,\mathrm{SED}}$ and find it slightly larger than what has been found in previous studies. Finally, we investigate which dust distribution reproduces better the estimated $A_{V,\mathrm{BD}}$ and $A_{V,\mathrm{SED}}$. We find that the attenuation towards old stars is consistent with the expectations for a standard galactic disc, where the stellar and dust distributions are mixed, while $A_{V, \mathrm{BD}}$ and the $A^\mathrm{young}_{V, \mathrm{SED}}$ are between the values expected for a foreground dust screen and the mixed configuration.}
   {}

   \keywords{galaxies: ISM - dust, extinction - ISM: atoms - ISM: molecules - ISM: abundances - galaxies: spiral}

   \maketitle

\section{Introduction}

Dust is a crucial component of the interstellar medium (ISM) and has a significant role in regulating the rate at which stars are formed, and hence galaxy evolution. Dust grains catalyse the transformation of atomic into molecular gas, which can cool and fragment forming stars under the power of gravity. Radiation from massive stars heats the gas but it is also absorbed by dust both locally, and in the diffuse ISM. The ultimate goal of understanding what drives star formation across a variety of galaxy discs, and consequently their evolution, can thus be achieved only by understanding the complex interplay among the various galaxy building blocks, such as cold and warm/hot ISM phases, the interstellar dust and the stellar populations.

   \begin{table*}
   \caption{The galaxies in our sample and their integrated properties.}
   \label{tab:sample}
   \tiny
   \begin{tabular}{c| c c|c| c c c| c c| c c | c c c }
\hline
\hline
   \multicolumn{1}{c|}{ID} &
   \multicolumn{1}{c}{RA} &
   \multicolumn{1}{c|}{DEC} &
   \multicolumn{1}{c|}{Type} &
   \multicolumn{1}{c}{D$_{25}$} &
   \multicolumn{1}{c}{D} &
   \multicolumn{1}{c|}{pw} &
   \multicolumn{1}{c}{$i$} &
   \multicolumn{1}{c|}{PA} &
   \multicolumn{1}{c}{$M_\mathrm{HI}$} &
   \multicolumn{1}{c|}{$M_\mathrm{H2}$} &
   \multicolumn{1}{c}{SFR} &
   \multicolumn{1}{c}{$M_\mathrm{star}$} &
   \multicolumn{1}{c}{$M_\mathrm{dust}$} \\
    & h : min : sec & º : $'$ : $''$ &  & arcmin & Mpc & kpc & º & º & 10$^9$ M$_\odot$ & 10$^9$ M$_\odot$ & M$_\odot$ yr$^{-1}$ & 10$^{9}$ M$_\odot$ & 10$^{7}$ M$_\odot$ \\
\textit{(1)} & \textit{(2)} & \textit{(3)} & \textit{(4)} & \textit{(5)} & \textit{(6)} & \textit{(7)} & \textit{(8)} & \textit{(9)} & \textit{(10)} & \textit{(11)} & \textit{(12)} & \textit{(13)} & \textit{(14)} \\
\hline
   NGC~0628 & 01:36:41.745 & +15:47:1.11 & Sc & 10.00 & 9.84 & 0.29 & 19.8 & 20 & 3.8 & 1.0 & 2.4$\pm$0.5 & 14.1$\pm$0.2 & 3.8$\pm$1.2\\
   NGC~3351 & 10:43:57.700 & +11:42:13.70 & SBb & 7.24 & 9.96 & 0.29 & 54.6 & 192 & 1.2 & 1.0 & 1.1$\pm$0.2 & 27.4$\pm$0.6 & 0.8$\pm$0.1\\
   NGC~3627 & 11:20:14.964 & +12:59:29.54 & SABb & 10.23 & 11.32 & 0.33 & 67.5 & 173 & 0.82 & 1.3 & 3.0$\pm$1.0 & 67.0$\pm$11.7 & 2.8$\pm$0.2\\
   NGC~4254 & 12:18:49.604 & +14:24:59.43 & Sc & 5.01 & 13.10 & 0.38 & 20.1 & 68 & 4.4 & 26.3& 5.2$\pm$0.5 & 13.4$\pm$0.2 & 2.2$\pm$0.1\\
   NGC~4321 & 12:22:54.831 & +15:49:18.54 & SABb & 6.17 & 15.21 & 0.44 & 23.4 & 30 & 2.9 & 20.9& 6.0$\pm$1.2 & 49.5$\pm$1.1 & 3.7$\pm$0.4\\
\hline
\hline
   \end{tabular}
\smallskip \phantom{}\\
  \footnotesize
   \noindent
   \textbf{Notes.}
   \textit{(1)} NGC  number; 
   \textit{(2) and (3)} J2000 Right Ascension and Declination;
   \textit{(4)} Hubble type;
   \textit{(5)} major axis of the optical 25 mag arcsec$^{-2}$ isophote;
   \textit{(6)} distance \citep{Anand2021}
   \textit{(7)} pixel width in physical scale;
   \textit{(8)} inclination;
   \textit{(9)} Position Angle \citep{Walter2008, Chung2009};
   \textit{(10)} total atomic gas mass  \citep{Walter2008, Chung2009};
   \textit{(11)} total molecular gas mass \citep{Leroy2008, Draine2007};
   \textit{(12)} global star-formation rate;
   \textit{(13)} total stellar mass;
   \textit{(14)} total dust mass. 
   When no reference is indicated, data is from the DustPedia archive.
\end{table*}

However, tracing the three main components of the ISM (atomic gas, molecules, dust) is not a simple task. The 21-cm line traces well the atomic hydrogen gas but it is challenging to detect individual galaxies at z~>~0.3 with current facilities \citep{Catinella2008, Verheijen+2010}. In contrast, CO lines can be observed up to intermediate redshifts, but they do not trace the dominant component of the molecular gas (H$_2$) and a conversion factor between the two is needed. Direct estimates of total gas mass of unresolved high-z galaxies are often not available or limited to the bright ones, because mm/radio spectroscopy is more time-consuming than optical spectroscopy. To overcome this limitation, optical spectra have been used to estimate gas masses \citep{Brinchmann2013}. Furthermore, the far-infrared (FIR) dust emission or the optical attenuation ($A_V$) inferred from SED-fitting ($A_{V,\mathrm{SED}}$) or from the Balmer decrement (BD, the ratio $F_{\mathrm{H}\alphaup}$/$F_{\mathrm{H}\betaup}$, usually converted to a dust attenuation $A_{V,\mathrm{BD}}$), have been used to trace the gas mass. A few studies have specifically explored how \textsc{Hi} and CO line emission, correlate with dust and $A_V$; and how dust emission correlates with $A_V$.

The literature does not yet provide a consistent picture regarding which gas phase shows a stronger correlation with dust. Despite it is more clear that the atomic gas has a loose correlation with the dust, it is still uncertain if dust traces better the molecular or the total gas. Those relations have been examined for various galaxy samples, both in global \citep[e.g.,][]{Corbelli2012, Grossi2016, Orellana2017, Casasola2020, Salvestrini2025} and in resolved scales \citep[e.g.,][]{Abdurrouf2022, Casasola2022}. 
\citet{Corbelli2012} studied 35 metal-rich galaxies in the Virgo Cluster and found that the dust mass correlates better with the total gas mass surface density ($\Sigma_\mathrm{gas}$), than with the atomic ($\Sigma_\mathrm{HI}$) or the molecular surface density ($\Sigma_\mathrm{H2}$). The latter result is confirmed by \citet{Orellana2017}, on 1,630 (z~<~0.1) galaxies, and \citet{Casasola2020}, on a sample of 252 DustPedia late-type galaxies.
For resolved galaxies such as NGC~5236 and NGC~0891, \citet{Foyle2012} and \citet{Hughes2014} found a weak correlation between $\Sigma_\mathrm{dust}$ and $\Sigma_\mathrm{HI}$, and a similarly tight relation between $\Sigma_\mathrm{dust}$--$\Sigma_\mathrm{H2}$ and $\Sigma_\mathrm{dust}$--$\Sigma_\mathrm{gas}$. This picture is confirmed by \citet{Casasola2022}, on a sample of 18 DustPedia galaxies at resolved (sub-kpc/kpc) scales.
Also \citet{Abdurrouf2022} for a sample of ten nearby galaxies (including four out of the five galaxies studied in this work) recover a tighter and more significant correlation for $\Sigma_\mathrm{dust}$--$\Sigma_\mathrm{gas}$, rather than for $\Sigma_\mathrm{dust}$--$\Sigma_\mathrm{H2}$, while for the $\Sigma_\mathrm{dust}$--$\Sigma_\mathrm{HI}$ relation the scatter increases and the correlation is moderate.

With respect to the relationship between the gas and $A_{V}$, for a sample of 222 local unresolved star-forming galaxies from the xCold GASS survey, \citet{Concas2019} find that  the BD can
be a powerful proxy of the molecular gas with a scatter of $\sim$0.3~dex, after correcting for disc inclination. They also find that the BD is not correlated with the \textsc{Hi} gas, while its correlation with the total gas is modest. The use of integrated quantities limits however the understanding of what drives the correlation and leaves open the question on the role of the \textsc{Hi} gas, more extended than the star forming disc and hard to estimate within the area traced by the BD if galaxies are unresolved. \citet{Barrera-Ballesteros2020}, find a tight correlation of the BD with the $\Sigma_\mathrm{gas}$, using only a low universal value for $\Sigma_\mathrm{HI}$ though. In that study they also find that although the scatter increases as smaller scales are sampled, the overall $\Sigma_\mathrm{dust}$--$\Sigma_\mathrm{gas}$ relation does not significantly depend on the scale. However, the question of whether the correlation of the BD with \textsc{Hi} or total gas is tighter than with the CO alone remains uncertain. On kpc scales, Concas et al. (in preparation) analyze ALMAQUEST data and  find a clear correlation between $A_{V,\mathrm{BD}}$ and $\Sigma_\mathrm{H2}$ (derived from CO(1-0) observations), with significantly smaller dispersion than  found by \citet{Barrera-Ballesteros2020}. Using the $A_{FUV}$, for a sample of 4 resolved and 27 unresolved galaxies, \citet{Boquien2013} found little correlation of the attenuation with the $\Sigma_\mathrm{HI}$, and a good correlation with $\Sigma_\mathrm{H2}$ and $\Sigma_\mathrm{gas}$.

Regarding the correlation between $\Sigma_\mathrm{dust}$ and the BD, \citet{Farley2025} examine how it varies on global scales, assuming several dust geometries and using a large sample from the Galaxy and Mass Assembly (GAMA) survey. In a study based on eight spatially resolved nearby galaxies, \citet{Kreckel2013} found a correlation between $\Sigma_\mathrm{dust}$ and $A_{V,\mathrm{BD}}$, however cautioning against using the BD to infer dust properties globally. 

In the current study, for the first time, we make a full comparison among dust surface density, $\Sigma_\mathrm{dust}$, surface density of all gas components, $\Sigma_\mathrm{HI}$, $\Sigma_\mathrm{H2}$, $\Sigma_\mathrm{gas}$ and optical attenuation derived from both BD, $A_{V,\mathrm{BD}}$, and SED fitting, $A_{V,\mathrm{SED}}$, relative to a sample of five nearby spiral galaxies observed with good spatial resolution. To achieve this, we perform a pixel-by-pixel SED-fitting analysis to derive  their physical properties (e.g. $\Sigma_\mathrm{dust}$ and $A_{V,\mathrm{SED}}$) in the lowest possible resolved scale. We create maps of the $A_{V,\mathrm{BD}}$ and $\Sigma_\mathrm{gas}$, extracted from H$_\alphaup$, H$_\betaup$ and CO, \textsc{Hi} maps, respectively. In Section \ref{sec:sample} we present the data, the data processing steps  and our methodology. Results are discussed in Section \ref{sec:discussion}. Finally, our findings are summarised in Section \ref{sec:sum}. In Appendix~\ref{ap:maps} we show spatially resolved maps of several physical quantities used or derived for each galaxy in the sample. In Appendix~\ref{ap:alpha_co} we discuss possible variations of the CO-to-H$_2$ conversion factor and of the $^{12}$CO J=2-1/J=1-0 line ratio. In Appendix~\ref{ap:atte} we present how attenuations are computed for simple geometries.

\section{Sample, data and processing}\label{sec:sample}

The main aim of our project is to examine a sample of large and well resolved local spiral galaxies with atomic and molecular gas maps, integral-field optical spectra and dust emission data, available throughout the star-forming disc. We selected all the nearby galaxies with the aforementioned observational data, which also have low or moderate inclination \mbox{($i$ < 70$^{\circ}$)} to avoid overestimating the $A_V$ \citep{Concas2019}. The five selected galaxies are NGC~0628, NGC~3351, NGC~3627, NGC~4254 and NGC~4321. Detailed information about integrated properties of our targets are shown in Table \ref{tab:sample}.

We used data from the DustPedia research project\footnote{http://dustpedia.astro.noa.gr} which includes far-ultraviolet (FUV) to sub-millimetre observations of 875 nearby galaxies that lie within 40~Mpc distance and have an optical diameter larger than 1 arcmin. All DustPedia galaxies are observed by the \textit{Herschel} Space Observatory \citep{Pilbratt2010} which helps constrain their dust content. For more information on the DustPedia project, we refer the reader to \citet{Davies2019} and \citet{Clark2018}. 

\subsection{The data}\label{sec:analysis}

To perform a spatially resolved analysis of the relation between the dust attenuation and the gas surface density, down to a sub-kpc physical scale (which corresponds to the pixel size of SPIRE-250~{\textmu}m image), we processed the data following specific steps in order to create a homogeneous dataset with common resolution and pixel scale across all wavebands. The pixel width of each galaxy, in physical units, is listed in Tab.~\ref{tab:sample}. The multi-wavelength data used in the current study are listed in Table \ref{tab:bands}. Below we describe the data processing.

\subsubsection{Galaxy images from FUV to FIR bands}\label{sec:steps}

   \begin{table}
   \caption{Summary of the multi-wavelength observational data used in the current analysis with their references.   }
   \label{tab:bands}
   \tiny
   \begin{tabular}{l c}
\hline
\hline
   \multicolumn{1}{l}{Survey/Telescope} &
   \multicolumn{1}{c}{Band name (effective wavelength; $\lambda$)} \\
\hline
   VLT-MUSE$^a$ & H$_\betaup$ (4861~\AA)\\
   VLT-MUSE$^a$ & H$_\alphaup$ (6562~\AA)\\
\hline   
   GALEX$^b$ & FUV (153~nm), NUV (227~nm)\\
   SDSS$^c$ & u (353~nm), g (475~nm), r (622~nm),\\
   ~ & i (763~nm), z (905~nm)\\
   2MASS$^d$ & J (1.24~{\textmu}m), H (1.66~{\textmu}m), Ks (2.16~{\textmu}m)\\
   WISE$^e$ & W1 (3.4~{\textmu}m), W2 (4.6~{\textmu}m), W3 (12~{\textmu}m), W4 (22~{\textmu}m)\\
   SPITZER$^f$ &  IRAC$^g$ (3.6~{\textmu}m, 4.5~{\textmu}m, 8.0~{\textmu}m, 12~{\textmu}m),\\
   ~ & MIPS$^h$ (24~{\textmu}m)\\
   HERSCHEL$^i$ & PACS$^j$ (70~{\textmu}m, 100~{\textmu}m, 160~{\textmu}m),\\
   ~ & SPIRE$^k$ (250~{\textmu}m)\\
\hline
   HERACLES$^l$ & $^{12}$CO(2-1) (1.3~mm)\\
   THINGS$^m$/VIVA$^n$ & \textsc{Hi} (21~cm)\\  
\hline
\hline
   \end{tabular}
   \smallskip \phantom{}\\
  \footnotesize
   \noindent
   \textbf{Notes.}
   $^a$\citet{Emsellem2022};
   $^b$\citet{Martin2005}; \citet{Morrissey2007};
   $^c$\citet{York2000}; \citet{Eisenstein2011};
   $^d$\citet{Skrutskie2006};
   $^e$\citet{Wright2010};
   $^f$\citet{Werner2004};
   $^g$\citet{Fazio2004};
   $^h$\citet{Rieke2004};
   $^i$\citet{Pilbratt2010};
   $^j$\citet{Poglitsch2010};
   $^k$\citet{Griffin2010};
   $^l$\citet{Leroy2009};
   $^m$\citet{Walter2008};
   $^n$\citet{Chung2009};
   \end{table}

We selected DustPedia images from FUV to 250~{\textmu}m emission and processed them following the steps described hereafter.
\begin{itemize}
\item[\textit{(i)}] We first removed foreground stars in the field of view of images up to the NIR, using the \citet{Cutri2003Cat} 2MASS all-sky catalogue of point sources. To optimally remove a foreground star, we assumed that the area to be removed depends on the brightness of the star and on the spatial resolution (FWHM) of the image. 
\item[\textit{(ii)}] Fluxes in bands shorter than 10~{\textmu}m were corrected for foreground galactic extinction, following the same methodology presented in \citet{Clark2018}. We assume R$_{\mathrm{V}}$=3.1 with a \citet{Cardelli1989} extinction curve. The A$_{\lambdaup}$ over E$(B-V)$ ratio is taken from \citet{GildePaz2007} for the FUV and NUV bands and from \citet{Schlafly&Finkbeiner2011} for the other bands (see the IRSA Dust Extinction Service\footnote{https://irsa.ipac.caltech.edu/applications/DUST/}).

\item[\textit{(iii)}] We convolved the data to the point spread function (PSF) of the band with the poorest resolution, which is the SPIRE-250~{\textmu}m band (18"). We used the \citet{Aniano2011} set of convolution kernels to degrade the resolution of all the shorter bands to the SPIRE-250~{\textmu}m PSF. In Sect.~\ref{sec:fitting} we comment on the impact of neglecting the longest wavelength (and poorest resolution) SPIRE bands.

\item[\textit{(iv)}] After convolving the
images, we used \textsc{iraf} \citep{Tody1986,Tody1993} to create a reference frame for each band, centred on the coordinates of the galaxy centre in the SPIRE-250~{\textmu}m image. We also selected the pixel-size of the reference frame to be equal to the pixel-size of SPIRE-250~{\textmu}m band (6") and the creation of the reference frame was done with the \textsc{mkpattern} package. Using the \textsc{wregister} package, we registered all images on the reference frame and we ended up having aligned images from different telescopes with a common PSF and pixel scale.

\item[\textit{(v)}]Once we had all images on the same grid, we subtracted the background noise. To successfully subtract inhomogeneities of the background, such as moir\'e patterns (GALEX, SDSS), foreground emission (e.g., sky brightness in NIR, Galactic cirrus in MIR and FIR), instrumental gradient (common in GALEX and Spitzer bands), which could affect our results, we modelled the complex sky with the \texttt{Background2D} class of the \texttt{Photutils} package in \texttt{AstroPy}. 

\item[\textit{(vi)}]Finally, pixels with low signal-to-noise ratio (S/N), which consequently do not have a clear physical connection with the galaxy, were discarded by applying a 3-$\sigmaup$ cut to the images. After applying the pixel filtering, we took into consideration in our SED-fitting analysis only those pixels featuring at least eleven bands above 3-$\sigmaup$, providing that SPIRE-250~{\textmu}m and at least one more \textit{Herschel} band are included.
\end{itemize}

The flux uncertainty of each pixel is estimated as the quadrature sum of the standard deviation of the flux in an annulus around the galaxy and the photometric calibration uncertainty of each band. The calibration uncertainties are taken from \citet{Clark2018}. We manually added an additional source of error, equal to 10\% of the flux, to account for uncertainties in the models used in the SED-fitting modelling as well as for unknown systematic errors in the photometry, as suggested by \citet{Noll2009} and as it is typically done in previous analyses \citep[e.g.,][among others]{Nersesian2019, Paspaliaris2021, Paspaliaris2023}. 

\subsubsection{Gas mass surface density maps}

To trace the atomic and molecular gas content of the galaxies we used \textsc{Hi} 21~cm and $^{12}$CO(2-1) line emission maps, respectively. The 21-cm line emission maps (“moment 0”; in Jy beam$^{-1}$ m s$^{-1}$) are drawn from "The \textsc{Hi} Nearby Galaxy Survey" \citep[THINGS;][]{Walter2008} for NGC~0628, NGC~3351 and NGC~3627, and the "VLA Imaging of Virgo in Atomic Gas" \citep[VIVA;][]{Chung2009} survey for NGC~4254 and NGC~4321. THINGS is undertaken at Very Large Array (VLA) of the National Radio Astronomy Observatory program that performed 21-cm \textsc{Hi} observations of 34 nearby galaxies. We used the ROBUST dataset, providing a close to uniform synthesized beam with FWHM~$\approx$~6" across the image. VIVA is an \textsc{Hi} imaging survey of 53 late-type Virgo cluster galaxies, at a resolution of ~15". The $^{12}$CO(2-1) line integrated intensity maps (in K~km~s$^{-1}$) are taken from “The HERA CO-Line Extragalactic Survey” \citep[HERACLES;][]{Leroy2009}. HERACLES provides $^{12}$CO(2-1) emission maps of 48 nearby galaxies, observed with the IRAM-30m telescope, at a resolution of 13.4". 

The PHANGS-ALMA survey \citep{Leroy2021} provides high quality CO(2-1) maps of 90 nearby galaxies, including the ones that we study in the current work. However, due to the limited field-of-view for each galaxy, we prefer to use the HERACLES maps. It has been reported that the HERACLES CO(2-1) map of NGC~3627 might suffer from calibration issues \citep[see,][]{denBrok2021, Leroy2021}. In a recent study, however, \citet{Kovacic2025} compare this map with a more recent ALMA map and find only minor differences, of the order of 0.06$^{+0.11}_{-0.26}$~dex.

The integrated \textsc{Hi} and CO maps were convolved, regridded and 3-$\sigmaup$ filtered \citep[see,][for detailed information on the data processing of THINGS, VIVA and HERACLES data, respectively]{Walter2008,Chung2009,Leroy2009}. We assumed that their original PSF is a Gaussian of the corresponding size. To perform the 3-$\sigmaup$ filtering, in THINGS maps we used the robust noise in one channel map, provided in Tab. 2 of \citet{Walter2008}, while a flux calibration uncertainty of the order of 5\% dominates the measurements of the flux densities. In the case of VIVA maps, we adopted the rms values measured by \citet{Chung2009} (see their Tab. 2) using the areas outside of the \textsc{Hi} emission. For the HERACLES maps of NGC~0628 and NGC~3351, we use the rms intensity level provided by \citet{Leroy2009} (see their Tab. 3). For the other galaxies we estimated  the rms using the areas outside the CO emission in the convolved and regridded moment-zero error-maps. We then used the 3-$\sigmaup$ filtered maps to derive gas mass surface density maps.

The atomic gas mass surface density, $\Sigma _{\mathrm{HI}}$, is calculated under the assumption of optically thin emission using the \textsc{Hi} 21-cm line intensity maps as
\begin{equation}
\Sigma_{\mathrm{HI}} \hspace{0.1cm} [\mathrm{M}_{\odot} \hspace{0.1cm} \mathrm{pc}^{-2}] = 0.020 \hspace{0.1cm} I_{21\mathrm{cm}}\hspace{0.1cm} [\mathrm{K} \hspace{0.1cm} \mathrm{km} \hspace{0.1cm} \mathrm{s}^{-1}],
\end{equation}
where the \textsc{Hi} column density has been multiplied by a factor 1.36 to account for heavy elements \citep{Leroy2012}. The maps were converted from units of [Jy beam$^{-1}$ m s$^{-1}$] to antenna temperature [K km s$^{-1}$] using Eq. (1) of \citet{Walter2008} and the synthesised beam size given in their Tab. 2 for THINGS galaxies, and in Tab. 2 of \citet{Chung2009} for VIVA galaxies.

The surface density of the molecular gas mass, $\Sigma _{{\mathrm{H2}}}$, is derived using the $^{12}$CO(2-1) line brightness, using the following equation \citep{Leroy2012},
\begin{equation}
\Sigma _{\mathrm{H2}} \hspace{0.1cm} [\mathrm{M}_{\odot} \hspace{0.1cm} \mathrm{pc}^{-2}] = 6.3 \hspace{0.1cm} I_{\mathrm{CO}} \hspace{0.1cm} [\mathrm{K} \hspace{0.1cm} \mathrm{km} \hspace{0.1cm} \mathrm{s}^{-1}],
\end{equation}
which assumes a CO(2-1)-to-CO(1-0) line ratio $R_{21}$ = 0.7 \citep[see e.g.,][]{Leroy2009, Schruba2011} and a Milky Way (MW) CO-to-H$_2$ conversion factor $\alphaup_{\mathrm{CO}}$ = 4.4 M$_{\odot}$ pc$^{-2}$ (K km s$^{-1}$)$^{-1}$  \citep{Bolatto2013} which corresponds to $X_{\mathrm{CO}}$ = 2 $\times$ 10$^{20}$ cm$^{-2}$ (K km s$^{-1}$)$^{-1}$ and includes the correction for heavy elements. The assumed $\alphaup_{\mathrm{CO}}$ is an intermediate value among those determined by various studies \citep[e.g.,][]{Strong1996, Dame2001, Abdo2010, Shetty2011}. This specific $\alphaup_{\mathrm{CO}}$ value is found to be optimal when investigating nearby spiral galaxies with metallicities close to solar \citep[e.g.,][]{Wong2002, Leroy2008}, such as our targets. Both $R_{21}$ and $\alphaup_{\mathrm{CO}}$ are assumed to be constant throughout galaxy disks unless stated differently. Possible variations of $\alphaup_{\mathrm{CO}}$ and $R_{21}$ are discussed in Ap.~\ref{ap:alpha_co}. The total gas mass surface density, $\Sigma _{\mathrm{gas}}$, is the sum of 
$\Sigma_{\mathrm{HI}}$ and
$\Sigma _{\mathrm{H2}}$.

\subsection{Dust-attenuation maps from Balmer decrement}

With the aim to estimate the dust absorption from the BD, we used data from the PHANGS-MUSE survey \citep{Emsellem2022}. PHANGS-MUSE used the MUSE integral field spectrograph at the VLT to map 19 star-forming disc galaxies, at a resolution lower than 1". For each galaxy, spectral cubes were created by mosaicking several pointings (up to 15) and convolving the final frames to a common angular resolution ($\sim$1") across all wavelengths.

\citet{Emsellem2022} processed the PHANGS-MUSE spectral cubes and provide, among several data products, the maps of emission in the  H$_\alphaup$ and H$_\betaup$ lines. We took those maps and convolved and regridded them to match the resolution of the rest of our data (18").  Pixels with \texttt{NaN} values were treated with interpolation during the convolution. However, if those pixels are at the edge of the map there are not enough nearby values and the interpolation might be unsuccessful. Such pixels are not considered in our analysis. We finally estimated the BD from the ratio of the convolved and regridded emission line maps. The approach of estimating the BD after smoothing the maps at the desired spatial resolution is physically accurate and in addition helps in recovering the signal in some pixel of the H$_\betaup$ maps. Being this line less bright than  H$_\alphaup$, by smoothing the maps we enhance its S/N and obtain a better coverage of the BD map.

The BD is usually converted into a V-band attenuation using
\begin{equation}
A_{V, \mathrm{BD}} = \frac{2.5\left[
\log_{10}(F_{\mathrm{H}\alphaup}/F_{\mathrm{H}\betaup}) - \log_{10}(F_{\mathrm{H}\alphaup}/F_{\mathrm{H}\betaup})_0\right]
}{k(\mathrm{H}\betaup) - k(\mathrm{H}\alphaup)},
\label{eq:avbd}
\end{equation}
where the term at the numerator is the difference between the attenuations in the two Balmer lines, $A_{\mathrm{H}\betaup, \mathrm{BD}}-A_{\mathrm{H}\alphaup, \mathrm{BD}}$, as derived from the observed BD and its intrinsic (i.e.\ unattenuated) value, $(F_{\mathrm{H}\alphaup}/F_{\mathrm{H}\betaup})_0$; that at the denominator is the difference between the attenuation laws (normalised to the V-band). 
We used $(F_{\mathrm{H}\alphaup}/F_{\mathrm{H}\betaup})_0$ = 2.86 as given by the assumption of a Case B recombination with a typical electron density of 10$^{-4}$~cm$^{-2}$ and a temperature of 10\,000~K, as for \textsc{Hii} regions \citep{Hummer1987, Osterbrock2006}, similarly to what has been done in previous studies \citep[e.g.][]{Kreckel2013, Momcheva2013, Piotrowska2020}. We also used $k(\mathrm{H}\alphaup)=0.77$ and $k(\mathrm{H}\betaup)=1.18$ from \cite{Kreckel2013}, which are very close to the values for the average ($R_V=3.1$) MW extinction law of \citet{Fitzpatrick1999}.

\subsection{Spectral energy distribution modelling}\label{sec:fitting}

We used the \texttt{CIGALE} SED-fitting code (see \citealt{Boquien2019}, and references therein) to model the SEDs of each pixel of the galaxies in our sample and retrieve maps of physical properties, such as stellar mass ($M_\text{star}$), dust mass ($M_\text{dust}$), star-formation rate (SFR) and $A_V$. Having defined a grid of values for the parameters of the various modules for the stellar, gas, and dust components and taking into account the dust attenuation, \texttt{CIGALE} creates a library of model SEDs which compares to the multi-wavelength observations in order to select the model SED that fits best to the data, through Bayesian inference. \texttt{CIGALE} includes all the different components in such a way that the amount of energy absorbed and re-emitted by the dust grains is fully conserved \citep{Noll2009, Roehlly2014}. 

   \begin{figure*}[t!]
   \centering
   \includegraphics[width=\textwidth]{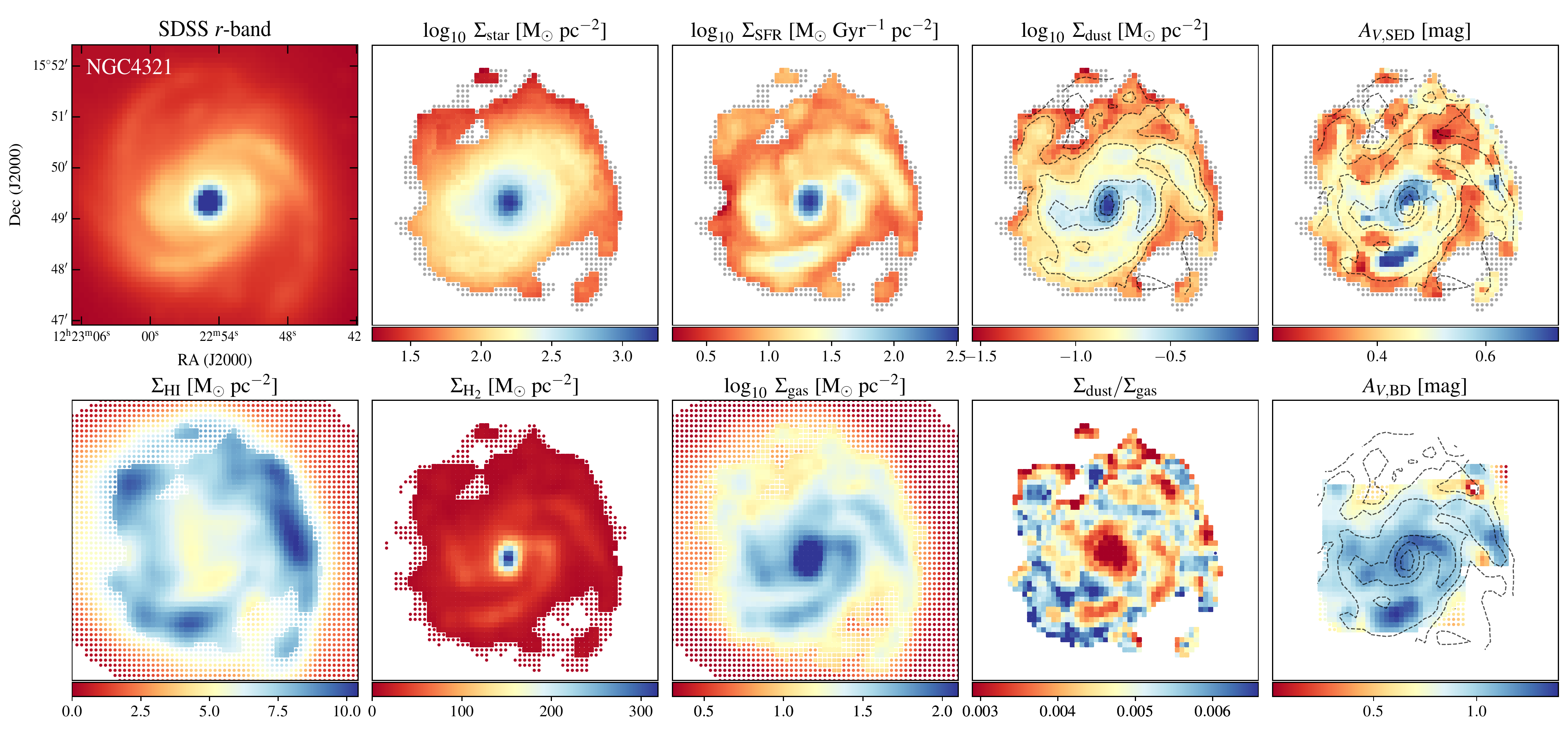}
   \caption{Maps of observed and derived properties of NGC~4321. \textit{Top}: Convolved and regridded SDSS r-band, logarithms of stellar mass surface density ($\Sigma_\mathrm{star}$), star-formation rate surface density ($\Sigma_\mathrm{SFR}$), dust mass surface density ($\Sigma_\mathrm{dust}$), as well as SED-fitting derived attenuation in the V-band ($A_{V,\mathrm{SED}}$), from left to right.
   \textit{Bottom}: Surface density of atomic gas mass ($\Sigma_\mathrm{HI}$), surface density of molecular gas mass ($\Sigma_\mathrm{H2}$), logarithm of the total gas mass surface density ($\Sigma_\mathrm{gas}$), dust-to-gas ratio and attenuation in the V-band as derived by the Balmer decrement ($A_{V,\mathrm{BD}}$), from left to right.
   All maps are at a resolution of 18".
   Gray points correspond to pixels that are rejected because they do not have a sufficient number of data points in the FIR regime (see Sec.~\ref{sec:steps}) or because they are unreliable (see Sec.~\ref{sec:fitting} for more details).
   The $\Sigma_\mathrm{HI}$ and $\Sigma_\mathrm{H2}$ maps extend up to their corresponding 3$\sigmaup$ limit. 
   In the $\Sigma_\mathrm{HI}$, $\Sigma_\mathrm{H2}$ and $A_{V,\mathrm{BD}}$ maps pixels that correspond to the areas that are excluded by the SED-fitting analysis are plotted with smaller dots. 
   In the $\log_{10}\Sigma_\mathrm{gas}$ map, pixels excluded by the SED-fitting analysis and having both \textsc{Hi} and CO detection are depicted by smaller squares, while pixels with only \textsc{Hi} and not CO are plotted with dots.
   In the dust-to-gas ratio maps we limit the colour-coding to the $5^{th}-95^{th}$ percentile range for illustrative purposes.
   Contours are taken from the $\log_{10}$~$\Sigma_\mathrm{dust}$ [M$_\odot$~pc$^{-2}$] maps with a lowest contour at -1.5 and linear spacing with the highest at 0.
   }
   \label{fig:mapsNGC4321t}
   \end{figure*}

Since we aim to investigate properties of a subset of galaxies that belong to the DustPedia sample, our parameter space is based on the parameter grid used in the reference DustPedia sample, introduced by \citet[see their Table 1]{Nersesian2019} which has also been applied successfully to other samples of local galaxies \citep[see e.g.][]{Paspaliaris2021, Paspaliaris2023}. A flexible SFH is used, allowing for a late instantaneous burst or quenching episode 200 Myr before the current moment (i.e. module \texttt{`sfhdelayedbq'}; \citealt{Ciesla2015}). The stellar population is thus separated in two components, an old one with age >~200Myr and a young population with age <~200 Myr. We assume an Initial Mass Function as given by \citet{Salpeter} between 0.1 and 100~M$_\odot$ and use the \citet{Bruzual&Charlot} stellar population model. In addition to the typical stellar metallicity value Z=0.02, often used, \citep[e.g.,][]{Boquien2019, Nersesian2019, Paspaliaris2021, Paspaliaris2023} we also considered Z=0.008, and allowed the fitting algorithm to determine the optimal stellar metallicity values between the two options. To model the gas metallicity we have used the radial metallicity gradients from O/H abundances, as derived by \citet{Brazzini2024} with the reliable direct method based on electron temperature measurements. These gradients for our sample are rather flat with central metallicities reaching the solar value. For each galaxy we have the sampled radial variations of the gas metallicity across the star forming disc and, depending on its gradient, we have established from 1 (flat gradient) up to 3 possible values of gas metallicities. These are all within the two metallicity values considered for the stellar population. The emission from the stellar components as well as from the ionised gas surrounding massive stars (i.e. nebular emission) are attenuated using the same power-law-modified starburst attenuation curve (i.e. module \texttt{`dustatt\_calzleit';} \citealt{Calzetti2000}, for 150nm~<~$\lambda$~<~2200nm; \citealt{Leitherer2002}, for 91.2nm~<~$\lambda$~<~150nm). The power-law slope is a free parameter modifying the attenuation curve for each fit. Following previous studies \citep[e.g.,][]{Noll2009,Boquien2012,Boquien2016,Nersesian2019} the amplitude of the UV bump is set to be zero. The UV emission in our sample is covered by just the two GALEX bands. To better constrain the bump, additional data and preferably NUV spectra \citep{Buat2012} are required, currently not available for our sample. However, we investigated how a non-zero UV bump amplitude might affect our results by running also CIGALE after setting the UV bump amplitude to the highest value suggested by \citet{Battisti2025} (that uses \textit{Swift}/UVOT NUV data).  We did not find sensible variations in the CIGALE results. The attenuation of each stellar population can be estimated separately and depends on two additional free parameters, the colour excess of the young stars and the attenuation-reduction factor for the old stars. The attenuation of the young stars is derived by their estimated colour excess, which is allowed to take 15 possible values varying between zero and one. The attenuation of the light from the old stellar population can be inferred through the reduction factor, which is multiplied to the attenuation of the young stars and is allowed to take three possible values (i.e. 0.25, 0.50, 0.75). The \textsc{THEMIS} model \citep{Jones2017} is used to account for dust emission.

Separate runs were performed for each galaxy in our sample. The number of free parameters used in the current analysis is 11 and the parameter space does not change from pixel to pixel within a galaxy. As discussed above, the only parameter whose range might vary from galaxy to galaxy is the gas metallicity. The total number of models produced for each galaxy varies depending on the number of possible values we set for each free parameter, ranging from 99\,792\,000 to 299\,376\,000. 
We would like to stress that in order to keep the maximum number of models at a tolerable level for our available computational power, in favour of setting metallicity as a free parameter, we reduced the possible values of some parameters, such as $i)$ the galaxy's e-folding time ($\tau_\mathrm{main}$ [Myr]) by removing two, 1700 and 3900, out of ten values $ii)$ the fraction of small hydrocarbon solids ($q_\mathrm{hac}$) by removing one (0.28) out of eleven values and $iii)$ the minimum interstellar radiation field ($U_\mathrm{min}$ [Habing]), for which we use 12 instead of 14 values, by removing 0.15 and by substituting 0.5, 0.8 by 0.6 and 1.2 by 1.0. The rest of the parameter grid is shown in detail in Tab. 1 of \citet{Nersesian2019}.

To exclude unreliable estimates of the properties we use a criterion that is based on the comparison of the best-model value (\texttt{best}) and the likelihood-weighted mean value (\texttt{bayes}) estimated by \texttt{CIGALE}, for each pixel. This has been also adopted in other recent studies \citep[e.g.][]{Buat2021, Mountrichas2022, Koutoulidis2022, Chakraborty2025}. Specifically, we consider in our analysis only pixels where $\frac{1}{5} \leq \frac{\mathrm{\texttt{best}}}{\mathrm{\texttt{bayes}}} \leq 5$, for the SED-fitting derived properties. The \texttt{bayes} value weights all models allowed by the parameter grid, with the best-fit model having the heaviest weight \citep{Boquien2019}. The weight is based on the likelihood, $e^{-\chi^2/2}$, associated with each model. A large difference between the two values (i.e. \texttt{best}, \texttt{bayes}) indicates that the fitting did not achieve a reliable estimation of the corresponding parameter. From this criterion, only 24 pixels (1\%) were rejected from NGC~0628, 58 pixels (2.7\%) from NGC~3351, 13 pixels (1.5\%) from NGC~3627, 3 pixels (0.6\%) from NGC~4254 and 57 pixels (3.5\%) were rejected from NGC~4321. 

   \begin{figure*}[ht!]
   \centering
   \includegraphics[width=\textwidth]{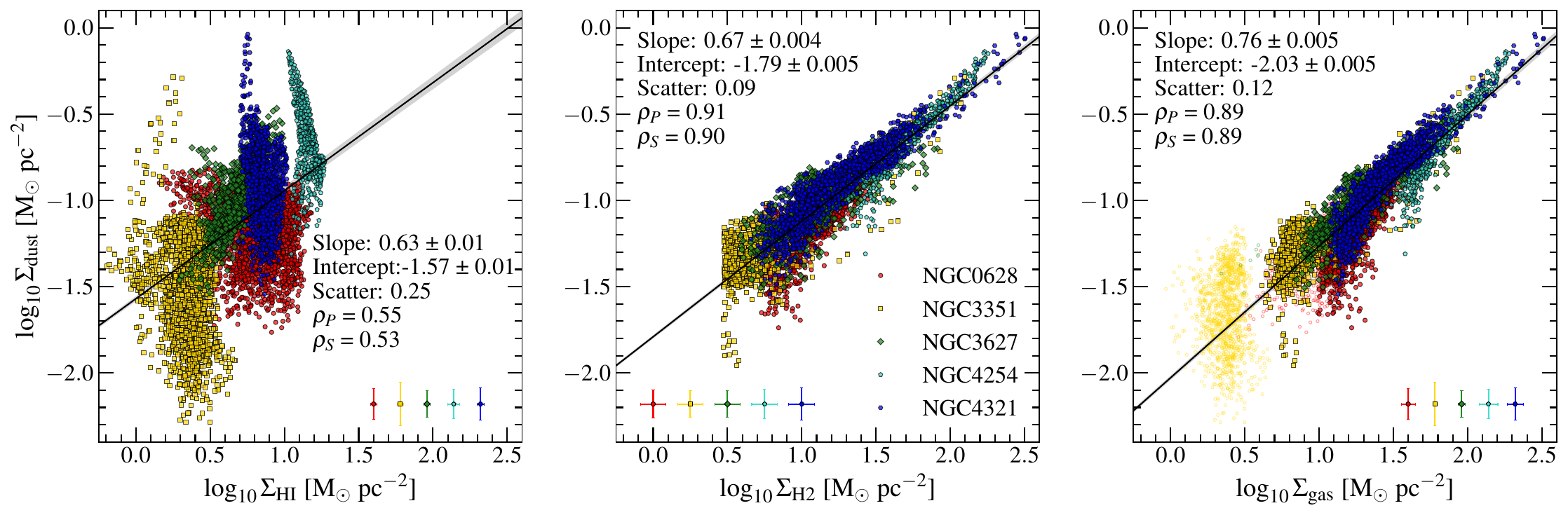}
   \caption{Resolved scaling relations of dust mass surface density with the atomic (\textit{left panel}), molecular (\textit{middle panel}) and total gas (\textit{right panel}) surface density. Each galaxy is represented by a different colour. The median uncertainties for each galaxy are shown in the lower-right or lower-left corner of the panels. Solid lines are the best linear fits to the full sample, while gray shaded area indicates the fit uncertainty. The slope, intercept, and scatter of each fit are also given along with the correlation coefficients. In the \textit{right panel} open circles show pixels where only \textsc{Hi} is detected.}
   \label{fig:dust-gas}
   \end{figure*}

As described in Sect.~\ref{sec:steps}, we discarded the SPIRE 350~{\textmu}m and 500~{\textmu}m bands in order to conduct the analysis at higher spatial resolution.
However, the choice has minimal consequence on the estimation of the dust luminosity and mass, as we found by performing three independent SED-fitting analyses for one of our targets, NGC~0628. Each time, we used different spectral coverage (i.e. up to 500~{\textmu}m, up to 350~{\textmu}m, and up to 250~{\textmu}m) and resolution (i.e. 36", 25", and 18", respectively) and created the radial profiles of $\Sigma_\mathrm{dust}$ using bin-widths equal to the resolution of each case. For all cases, we retrieved the same radial profile with the data lying within the uncertainties. We observed a mild overestimation of $\Sigma_\mathrm{dust}$ when we omitted the 500~{\textmu}m band, or both 500~{\textmu}m and 350~{\textmu}m bands, at a maximum of less than a factor of two, towards the very central part of the galaxy. In a recent study, \citet{Chastenet2024} also showed that the trends and parameter values are well reproduced without considering the two longer wavelength SPIRE bands.

A suite of maps of several parameters for NGC~4321 are shown in Fig.~\ref{fig:mapsNGC4321t}, as an example. In the top-left panel, we show the convolved and regridded SDSS r-band for a visual inspection of the morphological structure of the galaxy. The rest panels in the top row show maps of SED-fitting derived properties ($\log_{10}\Sigma_\mathrm{star}$, $\log_{10}\Sigma_\mathrm{SFR}$, $\log_{10}\Sigma_\mathrm{dust}$, $A_{V,\mathrm{SED}}$, from left to right). The bottom panels consist of maps of $\Sigma_\mathrm{HI}$, $\Sigma_\mathrm{H2}$, $\log_{10}\Sigma_\mathrm{gas}$, $\Sigma_\mathrm{dust}/\Sigma_\mathrm{gas}$ and $A_{V,\mathrm{BD}}$, from left to right. The corresponding maps of all galaxies in our sample can be found in Appendix~\ref{ap:maps}. 

\subsection{Data analysis: geometries and statistical methods}\label{sec:geom}

When comparing dust and gas surface densities, we consider face-on values, applying inclination corrections to each galaxy, under the assumption that the ISM is distributed in a thin disc. Attenuations may or may not be inclination dependent according to the geometrical distribution of the medium providing the attenuation in the case of the $A_{V,\mathrm{BD}}$, or to how dust is distributed with respect to the stellar component if we refer to the attenuated stellar light.  For the attenuation inferred by the BD, in the molecular phase for example, inclination corrections are less meaningful because molecular clouds have nearly spherical geometry. Consequently, in plotting the attenuation versus dust surface densities we refer to line of sight values (denoted by the LOS suffix) whether we consider dust mixed with the stars or distributed in a shell around star forming regions \citep[as also done by][]{Boquien2013, Kreckel2013}. For completeness, we briefly discuss the correlation coefficients for face-on values as well.

To estimate the scaling relations, we perform fitting through bayesian inference using the python UltraNest package \citep{Buchner2021}. UltraNest derives posterior probability distributions and the Bayesian evidence with the nested sampling Monte Carlo algorithm MLFriends \citep{Buchner2016,Buchner2019}. This method provides a robust estimate of the scaling relations between two quantities  incorporating the uncertainties of both quantities directly into the fitting process, leading to more reliable estimations of the fitted parameters (i.e., slope and intercept) and their uncertainties. For each correlation  we also provide the corresponding statistical coefficients that indicate its significance, i.e. Pearson's ($\rho_P$) and Spearman's ($\rho_S$) coefficients.

\section{Results and discussion}\label{sec:discussion}

The availability of a wide range of multiwavelength data for our galaxies allows us to examine the relation of dust in emission ($\Sigma_\mathrm{dust}$) with atomic gas traced by \textsc{Hi} ($\Sigma_\mathrm{HI}$), molecular gas traced by CO(2-1) ($\Sigma_\mathrm{H2}$) and the total gas ($\Sigma_\mathrm{gas}$), as well as the correlation of optical attenuation derived from the BD ($A_{V,\mathrm{BD}}$) and  the stellar light attenuation inferred by SED-fitting ($A_{V,\mathrm{SED}}$), with $\Sigma_\mathrm{gas}$ and $\Sigma_\mathrm{dust}$. Based on the $A_{V}$--$\Sigma_\mathrm{dust}$ relation, we investigate the dust spatial distribution using also predictions from simple radiative transfer models. Moreover, we compare $A_{V,\mathrm{BD}}$ to $A_{V,\mathrm{SED}}$. The correlations among these physical properties are explored in a resolved scale (346 pc, on average; see the pixel size of each galaxy in Tab.~\ref{tab:sample}).

The necessity of using only high S/N pixels for the derivation of the physical properties through the SED-fitting analysis, limits our study to the main part of the star-forming disc. Moreover, in the discussed correlations that include the $A_{V,\mathrm{BD}}$, we are slightly more limited to the inner part of the disc observed by MUSE. Thus, all the following correlations discussed here do not refer to the outskirts of the galaxies where the ISM is mostly atomic.

\subsection{Dust mass surface densities as gas tracers}\label{sec:dust-gas}

We examine here which gas phase establishes the best correlation with dust. In Fig.~\ref{fig:dust-gas} we present the relations between the dust and the gas mass surface densities for  galaxies in our sample. More specifically, in the left panel we show the $\log_{10}\Sigma_\mathrm{dust}$--$\log_{10}\Sigma_\mathrm{HI}$ relation, in the middle panel the relation between $\log_{10}\Sigma_\mathrm{dust}$ and $\log_{10}\Sigma_\mathrm{H2}$, while the correlation of the $\log_{10}\Sigma_\mathrm{dust}$ with the total gas, $\log_{10}\Sigma_\mathrm{gas}$, is displayed in the right panel.  

Although on average galaxies with higher $\Sigma_\mathrm{dust}$ have also higher $\Sigma_\mathrm{HI}$, the left panel of Fig.~\ref{fig:dust-gas} shows that these two quantities have a moderate correlation ($\rho_P$~=~0.55; $\rho_S$~=~0.53 in the log-log plane) because inside each galaxy $\Sigma_\mathrm{dust}$ is not often dependent on $\Sigma_\mathrm{HI}$. The average fitted slope is 0.63~$\pm$~0.01 and a significant scatter of 0.25~$\pm$~0.002 is present. An exception is NGC~3627 for which a moderate correlation is found on a resolved scale of 330 pc ($\rho_P$~=~0.67; $\rho_S$~=~0.66; slope:~0.94~$\pm$~0.04, scatter:~0.11).

The middle panel of Fig.~\ref{fig:dust-gas} shows instead that $\log_{10}\Sigma_\mathrm{dust}$ strongly correlates with $\log_{10}\Sigma_\mathrm{H2}$ having $\rho_P$~=~0.91 and $\rho_S$~=~0.90. The slope of the linear correlation is 0.67~$\pm$~0.004 with a scatter of only 0.09~$\pm$~0.001. The correlation holds also for individual galaxies for which slopes are similar to the global one. An increase in the scatter in the low $\Sigma_\mathrm{H2}$ regime, is a consequence of higher $\Sigma_\mathrm{dust}$ variations in galaxy's outskirts and in addition here the use of a constant $\alphaup_\mathrm{CO}$ might not be totally appropriate (see discussion in Ap.~\ref{ap:alpha_co}). The sublinear slope between $\log_{10}\Sigma_\mathrm{dust}$ and $\log_{10}\Sigma_\mathrm{H2}$ simply states that dust  distributions have shallower radial gradients than molecular gas distributions. Assuming that gas and dust can be described by radially exponential disks, it is straightforward to show that the correlation between $\Sigma_\mathrm{dust}$ and $\Sigma_\mathrm{H2}$ is the inverse of their scale-length ratios. From the slope of the linear fit in the $\log_{10}\Sigma_\mathrm{dust}$--$\log_{10}\Sigma_\mathrm{H2}$ relation, the dust scale-length is thus $\sim1.5$ times larger (on average) than the molecular gas scale-length. This is compatible with other results in the literature: for instance, \citet{Casasola2017} fit radial exponential profiles to $\Sigma_\mathrm{H2}$ and $\Sigma_\mathrm{dust}$ and find that, for their sample, the radial scale-length of the dust disc is on average about twice that of the molecular gas disc. We remind here that we have assumed constant values for the CO-to-H$_2$ conversion factors. If instead there are radial variations of $\alphaup_{\mathrm{CO}}$ (and/or $R_{21}$), the gradient of the molecular gas might be different from that inferred from the CO distribution.
Several studies have found that $\alphaup_{\mathrm{CO}}$ increases radially (see, e.g., \citealt{Bolatto2013}, \citealt{Chiang2024}, and Ap.~\ref{ap:alpha_co}). A radially increasing $\alphaup_{\mathrm{CO}}$ would give a flatter radial profile of $\Sigma_\mathrm{H2}$, closer to the $\Sigma_\mathrm{dust}$ gradient (in this case the slope of the $\Sigma_\mathrm{dust}$ and $\Sigma_\mathrm{H2}$ correlation  increases). For instance, using one of the radial dependent 
$\alphaup_{\mathrm{CO}}$ recipes suggested by \citealt{Chiang2024}, we found that the average slope of the correlation increases, from 0.67 to 0.77 (i.e. the dust scalelength is 1.3$\times$ that of molecular gas); however, there is no significant improvement in the correlation strength or in the scatter (see Fig.~\ref{fig:alphaCOs}).

A similar correlation, but with a steeper slope (0.76~$\pm$~0.005), is found between  $\log_{10}\Sigma_\mathrm{dust}$ and $\log_{10}\Sigma_\mathrm{gas}$, as shown in the right panel of Fig.~\ref{fig:dust-gas}. Here $\rho_P$~=~0.89, $\rho_S$~=~0.89, and a scatter of 0.12~$\pm$~0.001 are found, larger than for the $\log_{10}\Sigma_\mathrm{dust}$--$\log_{10}\Sigma_\mathrm{H2}$ relation. The correlation parameters remain unchanged if we omit pixels where only atomic gas is detected (open circles). The strength of the correlation between $\log_{10}\Sigma_\mathrm{dust}$ and $\log_{10}\Sigma_\mathrm{gas}$ (Fig.~\ref{fig:dust-gas}, right panel), is almost the same as that with the molecular gas only. This reflects the fact that the gas in the regions examined has a large contribution by the molecular phase. The steeper slope of the correlation is due to the increased contribution of atomic gas at large galactocentric distances. As for the relationship with the molecular gas only, the $\log_{10}\Sigma_\mathrm{dust}$--$\log_{10}\Sigma_\mathrm{gas}$ correlation seems to depend marginally on the properties of the sampled galaxies. This is in agreement with the findings of \citet{Abdurrouf2022} that find no significant galaxy-by-galaxy variations in $\Sigma_\mathrm{dust}$--$\Sigma_\mathrm{H2}$ and $\Sigma_\mathrm{dust}$--$\Sigma_\mathrm{gas}$ relations. By estimating the best-fit parameters for each galaxy individually, we notice that only the Virgo galaxies (NGC~4254, NGC~4321) have a steeper correlation (slope: 1.04~$\pm$~0.01, 0.84~$\pm$~0.01, respectively) than the mean slope. These findings hold also when we adopt a radially dependent $\alphaup_{\mathrm{CO}}$.

   \begin{figure}[t!]
   \centering
   \includegraphics[width=0.5\textwidth]{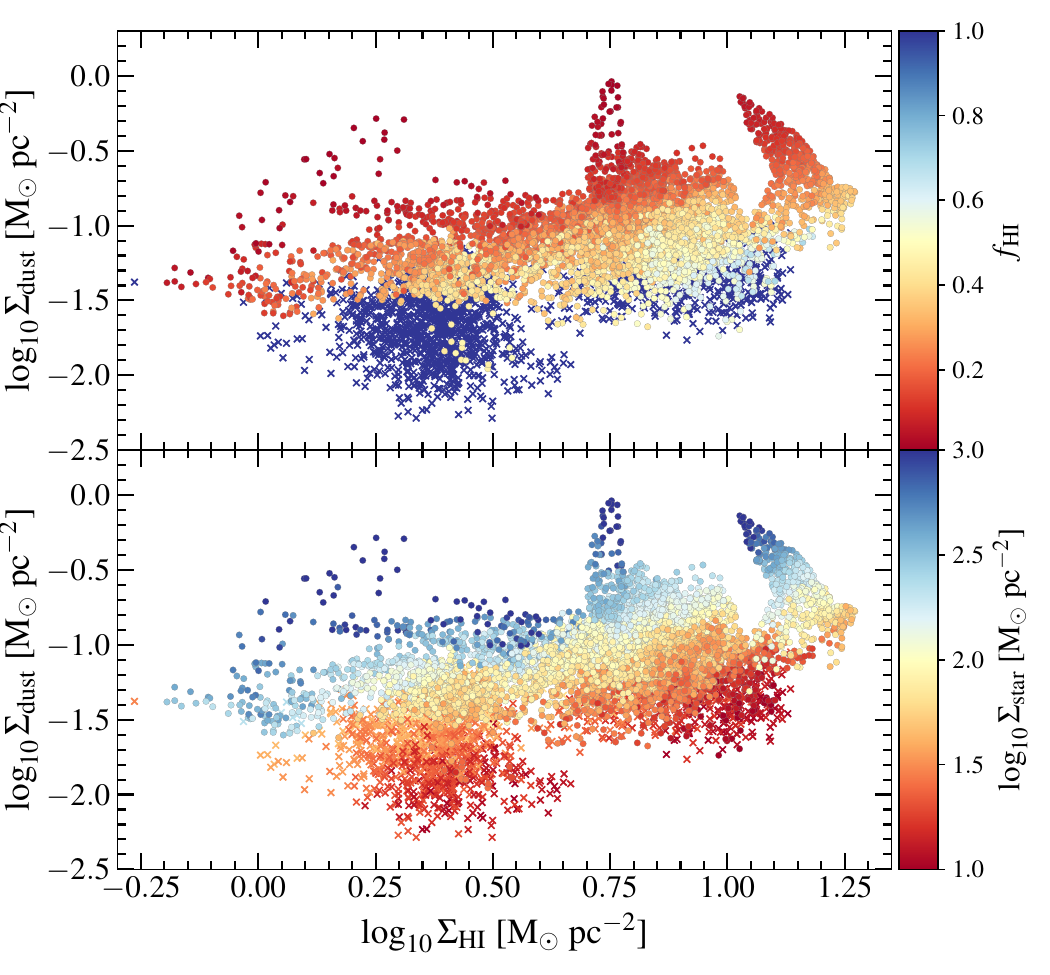}
   \caption{Dust mass surface density as a function of atomic gas mass surface density, colour-coded with atomic gas mass fraction (\textit{top-panel}) and stellar mass surface density (\textit{bottom panel}) for the whole sample. In both panels, crosses refer to pixels where only \textsc{Hi} gas is detected.
   }
   \label{fig:dg-CC}
   \end{figure}

   \begin{figure*}[t!]
   \centering
   \includegraphics[width=\textwidth]{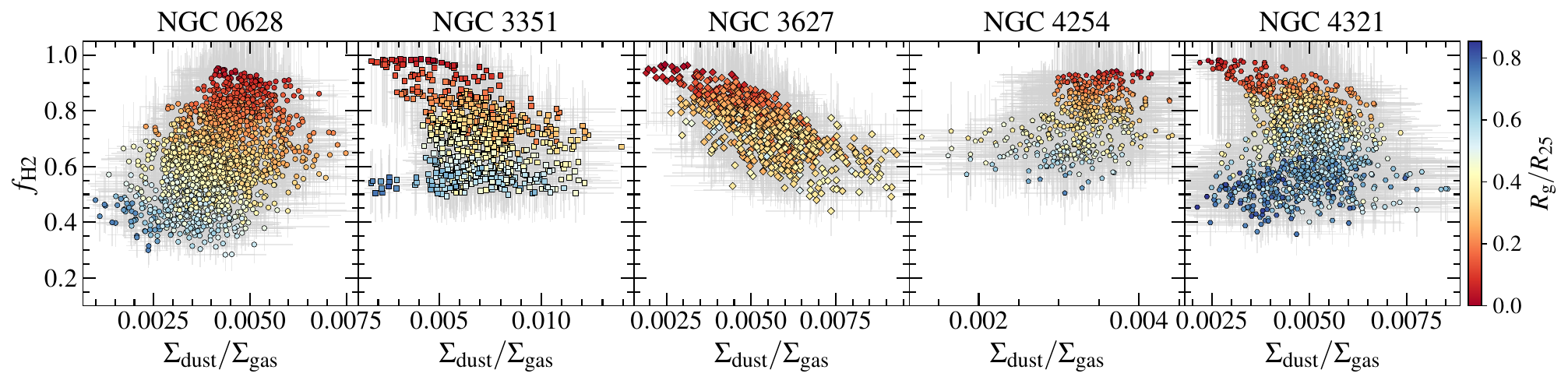}
   \caption{Molecular gas mass fraction as a function dust-to-gas mass ratio, for each galaxy in the sample. The data points are colour coded as a function of galactocentric radius in units of $R_{25}$. Uncertainties are plotted by gray error-bars.
   }
   \label{fig:fH2}
   \end{figure*}

\subsection{Molecular gas shielding layers and the dust-to-gas ratio}\label{sec:shield}

In the left panel of Fig.~\ref{fig:dust-gas} we notice that every galaxy has a narrow range of $\Sigma_\mathrm{HI}$; these vary only by a factor 2-3 across the disc extent we are examining, with the mean value for each galaxy that changes by about 0.8 dex across the sample.
What drives the changes of the mean \textsc{Hi} column density from one galaxy to another? One possibility is that variations of the mean $\Sigma_\mathrm{HI}$ from galaxy to galaxy are related to molecular shielding. Variations of the shielding \textsc{Hi} surface densities might be connected to variations of the dissociating radiation field, of volume gas densities or of the dust-to-gas mass ratio (DGR~=~$\Sigma_\mathrm{dust}/\Sigma_\mathrm{gas}$). For the DGR we expect $\Sigma_\mathrm{HI}$ to increase as the DGR decreases. The average DGRs found in the sampled regions are 0.0033~$\pm$~0.0005, 0.0041~$\pm$~0.001, 0.0048~$\pm$~0.0010, 0.0054~$\pm$~0.0013 and 0.0069~$\pm$~0.0020, for NGC~4254, NGC~0628, NGC~4321, NGC~3627 and NGC~3351, respectively. NGC~4254 is the richest galaxy in gas content and it shows the lowest DGR. Because of this low DGR, the galaxy needs the largest $\Sigma_\mathrm{HI}$ of the sample to shield the molecular phase. On the contrary, NGC~3351 has a low gas content and the highest DGR of our sample. Molecules in this case are shielded using a much lower \textsc{Hi} column density than for NGC~4254. 

The increase of the mean $\Sigma_\mathrm{dust}$ within the $\Sigma_\mathrm{HI}$ range for each galaxy might indicate an increase of $\Sigma_\mathrm{gas}$, i.e. an increase of the molecular fraction. To verify this in Fig.~\ref{fig:dg-CC} we plot the $\log_{10}\Sigma_\mathrm{dust}$ as a function of $\log_{10}\Sigma_\mathrm{HI}$, colour coded with atomic gas fraction ($f_\mathrm{HI}$~=~$M_\mathrm{HI}$/$M_\mathrm{gas}$) in the top-panel and with $\log_{10}\Sigma_\mathrm{star}$ in the bottom-panel. For each galaxy $f_\mathrm{HI}$ decreases going towards regions with higher dust, indicating an increase of the total gas column density, with the gas becoming more molecular. The bottom panel of Fig.~\ref{fig:dg-CC} underlines in fact that regions with higher $\Sigma_\mathrm{dust}$ are places where also $\Sigma_\mathrm{star}$ is higher, i.e. the gas layer has higher densities being more compressed by the local stellar gravity, and the formation of molecules is enhanced. 

We finally examine in Fig.~\ref{fig:fH2} how the molecular gas fraction ($f_\mathrm{H2}$~=~$M_\mathrm{H2}$/$M_\mathrm{gas}$) varies as a function of the DGR for each galaxy. We find that Sc galaxies (NGC~0628, NGC~4254) have a marginal positive trend with higher $f_\mathrm{H2}$ where the DGR is higher, while Sb galaxies hosting a bar (NGC~3351, NGC~3627) have a negative trend on average. A peculiar case is NGC~4321 (SABb) where a transition is observed with pixels having $f_\mathrm{H2}$~$\lesssim~$~0.8 (outer disc) exhibiting no correlation (or a positive correlation for $f_\mathrm{H2}$~$\lesssim~$~0.6), while pixels with higher $f_\mathrm{H2}$ (central area) having a negative correlation with the DGR. By visually inspecting the DGR map in Fig.~\ref{fig:mapsNGC4321t} we notice the low values in the central area and the rather inhomogeneous distribution in the rest of the disc. The scatter is generally large as in NGC~3351. Here, mainly the central pixels show some correlation with the DGR. 

Our findings might indicate a morphological evolution between Sc and Sb galaxies, possibly related to the presence of a bar. The bar might drag material (dust and gas) towards the centre of the galaxies; during this process, a significant part of the dust can be destroyed by the intense 
stellar radiation field in that area. This might explain why the DGR remains constant or decreases, while the molecular gas fraction can still increase towards the centre. This is apparent also in the DGR maps (see the corresponding maps in Fig.~\ref{fig:mapsNGC4321t} and Ap.~\ref{ap:maps}), where all barred galaxies exhibit a dust-poor centre. Another possibility is that central regions have a lower $\alphaup_{\mathrm{CO}}$ or a higher $R_{21}$, as we discuss  in detail in Ap.~\ref{ap:alpha_co}. For the rest of the discs, given the relatively flat metallicity gradients estimated by \citet{Brazzini2024}, with central metallicities of the order of solar, we don't expect strong variations of $\alphaup_\mathrm{CO}$. Although $\alphaup_\mathrm{CO}$ and R21 variations might give a somewhat different relation between $f_\mathrm{H2}$ and the DGR, our analysis in Ap.~\ref{ap:alpha_co} confirm the trends shown in Fig.~\ref{fig:fH2}. 

   \begin{figure*}[ht!]
   \centering
   \includegraphics[width=\textwidth]{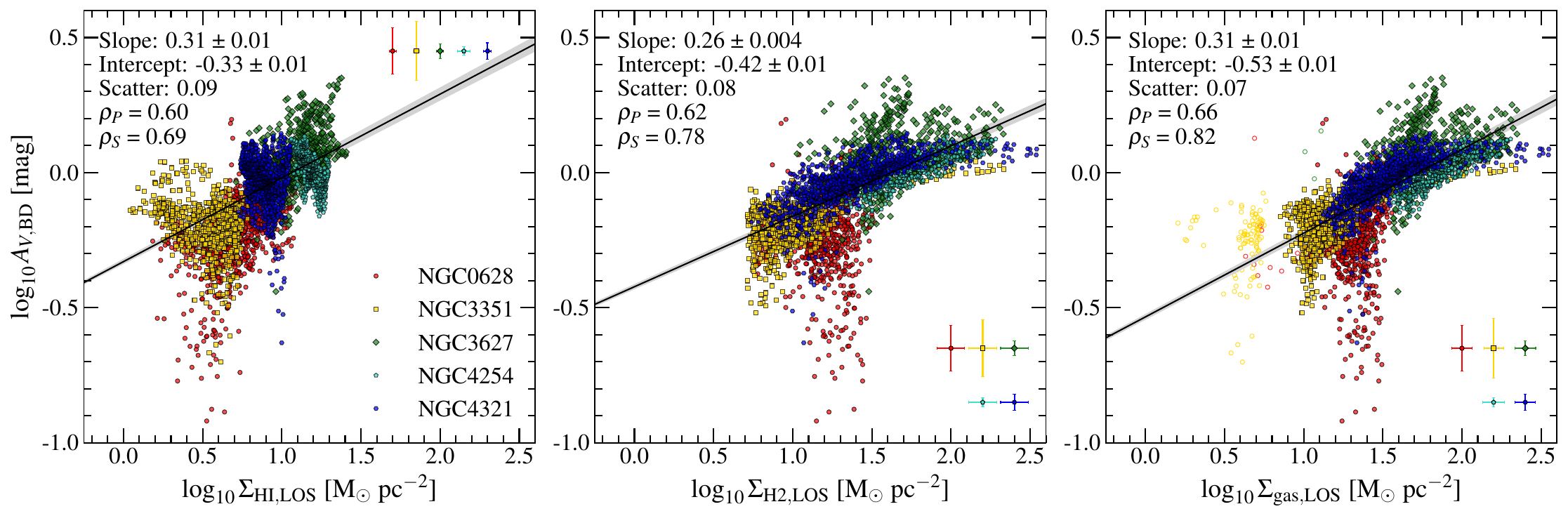}
   \caption{Attenuation on the V-band, derived from the Balmer decrement, as a function of the line-of-sight surface density of the atomic (\textit{left panel}), molecular (\textit{middle panel}) and total gas (\textit{right panel}) mass. Each galaxy is represented by a different colour. The median uncertainties for each galaxy are shown in the top- or the lower-right corner of the panels. Solid lines are the best linear fits to the full sample, while a shaded area indicates the fit uncertainty. The slope, intercept, and scatter of each fit are also given along with the correlation coefficients.}
   \label{fig:Av-SigmaGas}
   \end{figure*}

\subsection{The relations between $A_{V,\mathrm{BD}}$ and gas mass surface densities}

Here we examine the relation between the V-band attenuation as derived by the BD, $A_{V,\mathrm{BD}}$, with cold gas in the ISM. Our results are shown in Fig.~\ref{fig:Av-SigmaGas}, where  $\log_{10}A_{V,\mathrm{BD}}$ is plotted against  $\log_{10}\Sigma_\mathrm{HI}$ (left panel),  $\log_{10}\Sigma_\mathrm{H2}$ (middle panel) and  $\log_{10}\Sigma_\mathrm{gas}$ (right panel). 
The relations have similar scatter and flatter slopes (indicated in each panel) than those found between $\Sigma_\mathrm{dust}$ and $\Sigma_\mathrm{gas}$. Similarly to what we find for the $\log_{10}\Sigma_\mathrm{dust}$--$\log_{10}\Sigma_\mathrm{HI}$ relation, in the left panel of Fig.~\ref{fig:Av-SigmaGas} we see that galaxies which have higher $\Sigma_\mathrm{HI}$ have also on average higher $A_{V,\mathrm{BD}}$, but locally, in each galaxy, there is no correlation. The global correlation is moderate ($\rho_P$~=~0.60; $\rho_S$~=~0.69).

For $\log_{10}A_{V,\mathrm{BD}}$--$\log_{10}\Sigma_\mathrm{H2}$ relation, the positive correlation is clear both for individual galaxies and globally for the whole sample. The correlation is moderate if we rely on the correlation coefficients $\rho_P$~=~0.62, but the value of the Spearman coefficient $\rho_S$~=~0.78 indicates a stronger correlation of $A_{V,\mathrm{BD}}$ with the molecular gas surface density than with the atomic gas. The  $\log_{10}A_{V,\mathrm{BD}}$ has a similar correlation with the $\log_{10}\Sigma_\mathrm{gas}$ than with the $\log_{10}\Sigma_\mathrm{H2}$ ($\rho_P$~=~0.66; $\rho_S$~=~0.82). 

If we correct to face-on values both $A_{V,\mathrm{BD}}$ and gas surface densities, the above correlations weaken, the slopes are flatter and the scatter increases. As mentioned in Section~\ref{sec:geom}, corrections to face-on values are linked to the geometrical distribution of physical quantities which are not known a priori for the dust providing the attenuation. The stronger correlations we find for line-of-sight values compared to face-on values suggest that the geometric distribution of the absorbing medium is close to spherical. In this case also the attenuation linked to the atomic gas might be driven by HI envelopes around the star formation regimes rather than by the whole atomic gas disk layer. The correlations found between face-on cold gas mass surface densities and $A_{V,\mathrm{BD}}$ are also weaker than between the same gas surface densities and $\Sigma_\mathrm{dust}$, and underline that only part of the gas and dust mix, present in a galaxy, is responsible for the attenuation of stellar light.

We note that the fitting properties in the $\log_{10}A_{V,\mathrm{BD}}$--$\log_{10}\Sigma_\mathrm{gas}$ relations do not change if we exclude pixels detected only in \textsc{Hi} (open symbols). To ensure a robust estimation of the correlation, \citet{Barrera-Ballesteros2020} omit cases exhibiting $A_V$[mag]~<~0.2 thus excluding regions where CO emission likely drops rapidly due to the photo-dissociation of the CO molecules \citep{Dishoeck1988}. The exclusion of such pixels does not affect our findings concerning the estimated fitting parameters. In addition, motivated by the fact that we trace slightly different physical scales for galaxies in our sample, we examined the variation of the scatter with respect to the pixel width, but we find no trend. We underline that when we use a radially-dependent $\alphaup_{\mathrm{CO}}$ conversion as suggested by the \citet{Chiang2024} recipe, there is a marginal increase in the strength of the corresponding correlations, but the scatter remains similar (see Fig.~\ref{fig:alphaCOs}).

For NGC~3627 we find the highest $A_{V,\mathrm{BD}}$ values among all the regions studied in our sample (confirmed also by the $A_{V,\mathrm{SED}}$). Having this galaxy the highest inclination of all galaxies examined in this paper ($i$~=~67.5º) this finding is consistent with what \citet{Concas2019} have also found, namely  higher BDs in more inclined galaxies. However, some caution is needed for NGC~3627 because the high $A_{V}$ values are found mainly in the western arm, where an extended polarised radio ridge is, likely caused by ram-pressure compression \citep{Wezgowiec2012}. Ram pressure compression is able to increase locally gas and dust column densities, with a consequent increase of $A_{V}$  \citep[see e.g.,][]{Abramson2011}. Nonetheless, excluding NGC~3627 we find similar correlation coefficients for the line-of-sight $\log_{10}A_{V,\mathrm{BD}}$--$\log_{10}\Sigma_\mathrm{H2}$ relation ($\rho_P$~=~0.62; $\rho_S$~=~0.76) and again stronger correlation between $\log_{10}A_{V,\mathrm{BD}}$ and $\log_{10}\Sigma_\mathrm{gas}$ ($\rho_P$~=~0.66; $\rho_S$~=~0.82), as well as slopes and scatters similar to the ones found for the full sample.

   \begin{figure*}[ht!]
   \centering
   \includegraphics[width=\textwidth]{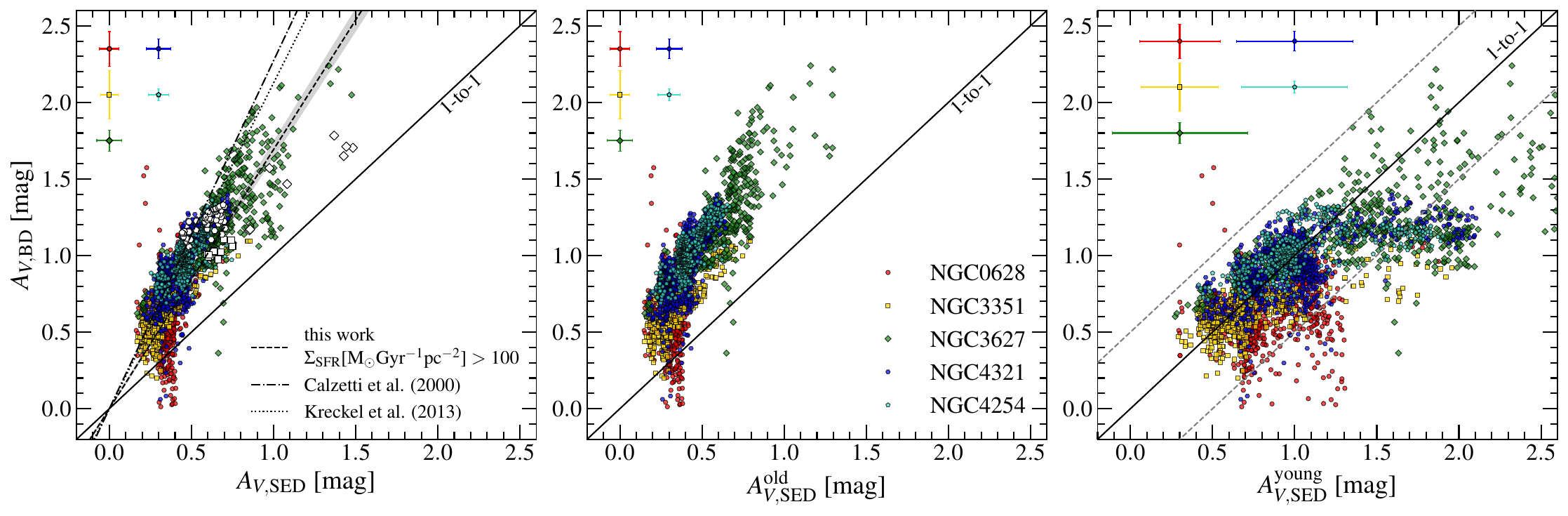}
   \caption{The correlation between the V-band attenuation derived by BD and SED-fitting (total, old and young stars, in \textit{left}, \textit{middle} and \textit{right panel, respectively}). Each galaxy is represented by a different colour. A solid line stands for the one-to-one relation. \textit{Left-panel}: $A_{V, \mathrm{BD}}$ versus the total attenuation estimated by CIGALE ($A_{V,\mathrm{SED}}$). Points with $\Sigma_\mathrm{SFR}$~[$\mathrm{M}_{\odot}~\mathrm{Gyr}^{-1}~\mathrm{pc}^{-2}$]~>~100 are shown in white. The best-linear fit to the latter points, assuming a zero intercept, is plotted with a dashed line and its uncertainty is indicated by the gray shaded area. A dash-dotted and a dotted line represent the corresponding results from \citet{Calzetti2000} and \citet{Kreckel2013}. \textit{Middle panel}: $A_{V, \mathrm{BD}}$ versus the attenuation of the old stellar component as estimated by CIGALE ($A^\mathrm{old}_{V, \mathrm{SED}}$). \textit{Right panel}: $A_{V, \mathrm{BD}}$ versus the attenuation of the young stellar component as estimated by CIGALE ($A^\mathrm{young}_{V, \mathrm{SED}}$). The plus-minus 0.5~dex area is defined by the dashed-gray lines.}
   \label{fig:Av-all}
   \end{figure*}

\subsection{Comparison of $A_{V,\mathrm{BD}}$ and $A_{V,\mathrm{SED}}$}\label{sec:bd-sed}

A visual inspection of the $A_V$ maps displayed in  Fig.~\ref{fig:mapsNGC4321t} and Ap.~\ref{ap:maps} shows that the attenuations computed via BD and SED-fitting have a similar pattern across the discs, although $A_{V, \mathrm{BD}}$ is systematically higher than the $A_{V, \mathrm{SED}}$. This is also shown in the left panel of Fig.~\ref{fig:Av-all}, where the $A_{V}$ derived from the two independent methods are compared.
If we focus on the star-forming regions with $\Sigma_\mathrm{SFR}$~[$\mathrm{M}_{\odot}~\mathrm{Gyr}^{-1}~\mathrm{pc}^{-2}$]~>~100 (white data-points) the dispersion is reduced. By performing a linear best-fitting analysis on the star-forming regions we find that
\begin{equation*}
    A_{V, \mathrm{SED}} = (0.59~\pm~0.05)~A_{V, \mathrm{BD}}.
\end{equation*}
The slope of our best fitted linear relation, shown with a dashed line, is slightly higher than the value found for the relation between emission line and stellar continuum  $A_{V}$ by \citet{Calzetti2000} (dash-dotted line, slope 0.44) and \citet{Kreckel2013} (dotted line, slope 0.47).
We underline that this result is sensitive to the selected attenuation law (i.e.\ the $k(\mathrm{H}\alphaup)$,
and $k(\mathrm{H}\betaup)$ values used in Eq.~\ref{eq:avbd}). 
In principle, the attenuation law found for each pixel by our SED-fitting analysis, under the assumption of a modified power-law starburst attenuation curve (see Sect.~\ref{sec:fitting}), might result in a different conversion factor between $F_{\mathrm{H}\alphaup}/F_{\mathrm{H}\betaup}$ and $A_{V, \mathrm{BD}}$. Nevertheless, we found that, when averaging over all pixels, that factor is very close to that adopted by \citet{Kreckel2013} and thus we use a uniform conversion factor.

The selected SFH and attenuation modules used in our SED-fitting analysis allow us to estimate separately the attenuation on the emission of the old and the young stars. In Fig.~\ref{fig:Av-all} middle and right panels, we compare $A_{V, \mathrm{BD}}$ with the attenuation on the old ($A_{V, \mathrm{SED}}^\mathrm{old}$) and young stars ($A_{V, \mathrm{SED}}^\mathrm{young}$), respectively. We find that the total attenuation estimated by \texttt{CIGALE} is dominated by $A_{V, \mathrm{SED}}^\mathrm{old}$, which is expected, since the bulk of the total stellar content in spiral galaxies is dominated by old stars \citep[see e.g.,][]{Nersesian2019, Paspaliaris2023}. The $A_{V, \mathrm{SED}}^\mathrm{young}$ though, despite the larger uncertainties, is in a quite good agreement with the $A_{V, \mathrm{BD}}$, with the bulk of the data-points following the one-to-one relation and 88.6\% of them lying within 0.5~dex (gray dashed lines). This indicates that \texttt{CIGALE} is able to estimate the attenuation in star-forming regions on resolved scales, using broad-band photometric data only. In a future study, it would be interesting to investigate if the scatter and the uncertainties might decrease by enriching the relative free parameters in the SED-fitting with more values and/or by including also narrowband photometric data (e.g. H$_\alphaup$ maps) that are available for all the galaxies in our sample. A plateau found at $A_{V,\mathrm{BD}}$~$\sim$~1.3~mag, might be attributed to the fact that $A_{V, \mathrm{BD}}$ and $A_{V, \mathrm{SED}}^\mathrm{young}$ trace different environments with different optical depths at high attenuation values (see also Sec.~\ref{sec:avsd} where, at high attenuations, $A_{V, \mathrm{SED}}^\mathrm{young}$ is more closely connected with a dust screen configuration than the $A_{V, \mathrm{BD}}$).

\subsection{The correlation of dust in absorption and dust in emission} \label{sec:avsd}

\begin{figure*}[ht!]
   \centering
   \includegraphics[width=\textwidth]{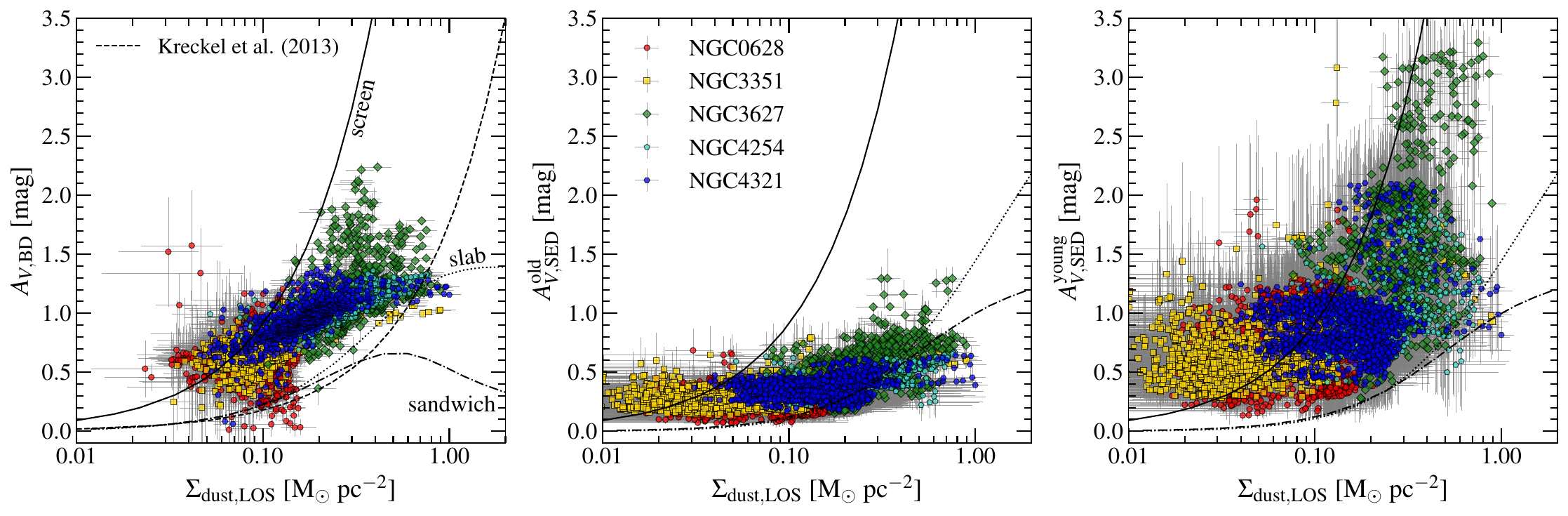}
   \caption{Relation of line-of-sight dust mass surface density with V-band attenuation derived by the BD (\textit{left panel}), attenuation of the old stars by the SED-fitting (\textit{middle panel}) and attenuation of the young stars also derived by \texttt{CIGALE} (\textit{right panel}). Uncertainties in both axes are shown in gray. The dust mass surface density estimated under the assumption of a foreground {\em screen} dust model is plotted with a solid line. The dotted and dash-dotted curves show the results of the {\em slab} and {\em sandwich} models, respectively: those in the \textit{left panel} are for the $A_{V, \mathrm{BD}}$--$\Sigma_\mathrm{dust}$ relation, those in the \textit{central} and \textit{right panel} for $A_{V}$--$\Sigma_\mathrm{dust}$.
    The best fit of \citet{Kreckel2013} is plotted by a dashed line in the \textit{left panel}.
   }
   \label{fig:Av-Dust}
   \end{figure*}

A visual comparison of the spatial distribution of $\Sigma_\mathrm{dust}$, $A_{V, \mathrm{BD}}$ and $A_{V, \mathrm{SED}}$ across galaxies' discs, can be done using the maps of Fig.~\ref{fig:mapsNGC4321t} and Ap.~\ref{ap:maps}, where contours of  $\Sigma_\mathrm{dust}$ are over-plotted in all three maps. The pixel-by-pixel relation for each galaxy is shown in Fig.~\ref{fig:Av-Dust} using the $A_{V, \mathrm{BD}}$ (left panel), the $A^\mathrm{old}_{V, \mathrm{SED}}$ (middle panel) and the $A^\mathrm{young}_{V, \mathrm{SED}}$ (right panel). The number of data-points is smaller in the first panel because of the restricted extent of MUSE data, compared to the \texttt{CIGALE} SED-fitted area. We observe an increasing trend of $\Sigma_\mathrm{dust}$ with $A_{V, \mathrm{BD}}$ and $A^\mathrm{young}_{V, \mathrm{SED}}$, rather than $A^\mathrm{old}_{V, \mathrm{SED}}$ that increases milder as a function of $\Sigma_\mathrm{dust}$. This suggests that nebular lines and young stars are more sensitive tracers of the dust mass surface density in our sample. Our finding is in agreement with the results of \citet{Munoz2009} who used the total IR-to-UV ratio, tracing intermediate age stars, to measure the attenuation, and \textit{Spitzer} imaging to estimate the radial dust properties of nearby galaxies. We also find our results in agreement with those of \citet{Kreckel2013} whose method of analysis is more similar to that adopted in the current study. 

The quantity $A_V$ is proportional to the dust surface density only when dust is distributed in a screen in front of the sources.
$\Sigma_\mathrm{dust}$ can be directly converted into $A_{V}$ after assuming a DGR (see Eq.~4 in \citealt{Kreckel2013}). This foreground {\em screen} model is shown in Fig.~\ref{fig:Av-Dust}, where we use the DGR of the THEMIS model, equal to 0.0074.
For the same surface density, lower extinctions can be obtained if dust and stars are distributed in the same layer (the {\em slab} model; \citealt{Disney1989,Calzetti1994}); and even lower
if the dust layer is internal to the stellar one
(the {\em sandwich} model; \citealt{Disney1989}), as emerges from radiative transfer analyses of edge-on galaxies \citep{Xilouris1999,Bianchi2007,DeGeyter2013}. Results from the 
\textit{slab} and {\em sandwich} models are shown in Fig.~\ref{fig:Av-Dust}. Rather than the analytical solution for a non-scattering medium \citep{Disney1989} or approximations to include scattering \citep{Calzetti1994,Kreckel2013,Boquien2013}
we made a full radiative transfer calculation using the TRADING code \citep{Bianchi2008}, assuming that the thickness of the dust layer is the same (for the {\em slab}) or half (for the {\em sandwich}) that of stars, and adopting the average properties of the THEMIS model (see Ap.~\ref{ap:atte} for details). $A_{V,\mathrm{BD}}$ for the {\em slab} and {\em sandwich} models, derived using Eq.~\ref{eq:avbd}, are shown by the dotted and dash-dotted lines of Fig.~\ref{fig:Av-Dust} (left panel), respectively. 
It is worth noting that $A_{V,\mathrm{BD}}$ is not a direct measure of $A_{V}$ and in particular the two quantities diverge for high dust column densities (optical depths). We plot $A_{V}$ for our models in the central and right panel of Fig.~\ref{fig:Av-Dust}. The difference is due to the fact that Eq.~\ref{eq:avbd} is formally correct only when the effective attenuations driving the difference between the observed and intrinsic $F_{\mathrm{H}\alphaup}/F_{\mathrm{H}\betaup}$
follow the same attenuation law, described by the $k(\mathrm{H}\alphaup)$ and $k(\mathrm{H}\betaup)$ values used in the formula. Instead, constant values are used in Eq.~\ref{eq:avbd}, while the true attenuation laws change with dust column density because of radiative transfer effects. This is true not only for the models shown in the figure, but also for simpler cases with no scattering (see Ap.~\ref{ap:atte}). However, for the surface density range shared by most pixels, $A_{V}$ and $A_{V,\mathrm{BD}}$ have similar trends and can be considered as lower limits to the V-band attenuation for our pixels. We also derived attenuations for an exponential model with parameters similar to those derived by fits of edge-on galaxies, but found little differences with the {\em sandwich} model (not shown here).

The left panel of Fig.~\ref{fig:Av-Dust} shows that $A_{V,\mathrm{BD}}$ is close to the {\em screen} model for low $\Sigma_\mathrm{dust}$ pixels, while it lies between the {\em screen} and the {\em slab} at higher surface densities.
This suggests that the dust that obscures the emitters in the star-forming areas are distributed in a mixed dust and emitting material, as well as a foreground {\em screen} dust component. 
\citet{Kreckel2013} reached similar conclusions, though their best fit to the $A_{V,\mathrm{BD}}$~--~$\Sigma_\mathrm{dust}$ correlation (dashed line) tend to be shifted towards higher surface densities; beside differences in the sample and analysis, this could be in part due to  different dust emission templates used to derive $\Sigma_\mathrm{dust}$ (they use those of \citealt{Draine2007}, which results in dust masses larger by about 50\% with respect to those derived using THEMIS; see \citealt{Nersesian2019}). Our findings are also in accordance with a recent study, in global scales, using GAMA data (i.e. BD and $\Sigma_\mathrm{dust}$), by \citet{Farley2025}. In that paper it is found that in the case of low dust content, the foreground {\em screen} and the mixed geometry (mentioned as \textit{distributed} in their study) both fall in the low BD region, as neither is able to provide enough attenuation to explain the high BD measurements. Furthermore, they report two distinct cases where, for high dust content, the dust in the {\em screen} model is optically thick and prevents emission lines  to escape, leading to a lack of measurements in that regime. In the case of the mixed geometry, lines originating from stars deeply embedded in dust clouds escape less (i.e. are more attenuated) than the ones in the outer regions, allowing for low BD measurements despite the high $\Sigma_\mathrm{dust}$ in these regions.

As shown also in the previous Section, $A^\mathrm{young}_{V, \mathrm{SED}}$ (right panel of Fig.~\ref{fig:Av-Dust}) follows a trend similar to that of $A_{V,\mathrm{BD}}$. Despite the large uncertainties, the attenuation estimated for young stars is consistent with that estimated from the BD.
Instead, $A^\mathrm{old}_{V, \mathrm{SED}}$ 
follows more closely the models where stars and dust are mixed (middle panel; {\em slab} and {\em sandwich} are very close for low column densities) suggesting that the radiation from the old stars is being obscured mainly by the diffuse star/dust geometry of a galactic disc. $A^\mathrm{old}_{V, \mathrm{SED}}$ also shows that the attenuation does not tend to zero at the lowest column densities, but is biased towards a non-null value. \citet{Boquien2013} noted a similar trend, with non-null attenuation at the lowest gas surface densities, and imputed it to the fact that attenuation is a luminosity-weighted quantity while surface densities are mass-weighted means over the finite extent of a resolution element. It is unclear whether this consideration applies also for our case, given the fact that the $\Sigma_\mathrm{dust}$ is derived from SED fits to FIR luminosity, unless the biasing due to luminosity weighting acts in different ways on the determination of $A^\mathrm{old}_{V, \mathrm{SED}}$ and $\Sigma_\mathrm{dust}$. Investigating this issue would require a multi-wavelength modelling of a galaxy and the fitting of its SED at different resolutions, which is beyond the scope of the present study.

However, correlations between $\Sigma_\mathrm{dust}$ and $A_{V,\mathrm{SED}}$ derived by the SED-fitting should be interpreted with some caution. The two properties are estimated by the same SED-fitting run and under the assumption of energy balance. Although, despite they are expected to be coupled, they are not tightly constrained by each other alone. For instance, a certain IR luminosity can be produced by different amounts of dust mass, under different heating conditions, which are not tightly constrained by the UV–optical data  used to fit the attenuation curve. Alternatively, higher $U_\mathrm{min}$ (i.e. higher dust temperature) can reduce the need for a high dust mass to explain the IR flux, while still producing a similar attenuation. Indeed, despite the caveat on energy balance our findings regarding the correlations between $A^\mathrm{old}_{V, \mathrm{SED}}$, $A^\mathrm{young}_{V, \mathrm{SED}}$ and $\Sigma_\mathrm{dust}$ are in accordance with the picture derived by the RT models.

\section{Summary and conclusions}\label{sec:sum}

We investigate the relation among the dust mass surface density ($\Sigma_\mathrm{dust}$), the gas mass surface density ($\Sigma_\mathrm{HI}$, $\Sigma_\mathrm{H2}$, $\Sigma_\mathrm{gas}$) as well as the optical attenuation ($A_{V, \mathrm{BD}}$, $A_{V, \mathrm{SED}}$), throughout the disc of a sample of five nearby spiral galaxies, down to sub-kpc scales. Our main results are summarised as follows:

\begin{itemize}
    \item In the regions sampled in this study, which have molecular-fraction $f_\mathrm{H2}\gtrsim0.3$, we find that $\Sigma _\mathrm{dust}$ and $A_{V,\mathrm{BD}}$ trace better molecular and total gas content, rather than  atomic gas. The correlations of $\Sigma _\mathrm{dust}$ with $\Sigma_\mathrm{H2}$ or  with $\Sigma_\mathrm{gas}$ have the strongest significance and hold for individual galaxies as well as for the whole sample.

    \item The scaling relations are marginally dependent on galaxy properties. The mean dust-to-gas mass ratio varies across our sample with each galaxy showing local enhancements. Molecular fractions increase as $\Sigma _\mathrm{dust}$ and $\Sigma _\mathrm{star}$ increase.

    \item The atomic gas mass surface density for each galaxy is in a restricted range which varies from galaxy to galaxy and is related to the column density that the gas needs to shield the molecular phase. Galaxies with a higher dust-to-gas mass ratio need a lower column density of \textsc{Hi} to provide the shielding, with local variations related to the $\Sigma _\mathrm{star}$. 
    
    \item The $A_{V,\mathrm{BD}}$ and the $A^\mathrm{young}_{V, \mathrm{SED}}$ that we estimated through SED-fitting are in a very good agreement, with 88.6\% of our data points differing by less than 0.5~dex. $A^\mathrm{young}_{V, \mathrm{SED}}$ inferred by \texttt{CIGALE} could serve as a tracer of the attenuation in larger samples, where IFU observations might be more expensive and time-consuming.

    \item We estimate the ratio of the $A_{V,\mathrm{BD}}$ over the total stellar $A_{V,\mathrm{SED}}$, for star-forming regions, equal to 0.59~$\pm$~0.05, slightly larger than the one found in previous studies. 

    \item Both $A_{V, \mathrm{BD}}$ and $A^\mathrm{young}_{V, \mathrm{SED}}$ are affected by a combination of a screen dust component and a mixed dust and emitting material. On the contrary, $A^\mathrm{old}_{V, \mathrm{SED}}$ as a function of the $\Sigma _\mathrm{dust}$ follows the 
    models where sources and dust are mixed (either the {\em slab} or the {\em sandwich} model).
    
\end{itemize}

\section*{Data availability}

Maps of the properties derived in the current study are only available in electronic form at the CDS via anonymous ftp to \url{cdsarc.u-strasbg.fr} (130.79.128.5) or via \url{http://cdsweb.u-strasbg.fr/cgi-bin/qcat?J/A+A/}.

\begin{acknowledgements}
We are grateful to the anonymous referee for a constructive report that improved the quality of the manuscript. We would like to thank Adam Leroy and Francesco Belfiore for productive discussions on this project and on the use of PHANGS-MUSE data.
We acknowledge financial support from: PRIN MIUR 2017 – 20173ML3WW\_001;  INAF mainstream 2018 program “Gas-DustPedia: A definitive view of the ISM
in the Local Universe”; INAF-Mini Grant 2024 program "Dust emission and optical extinction as gas tracers in star forming galaxies".
DustPedia is a collaborative focused research project supported by the European Union under the Seventh Framework Programme (2007–2013) call (proposal no. 606824). The participating institutions are: Cardif University, UK; National Observatory of Athens, Greece; Ghent University, Belgium; Université Paris-Sud, France; National Institute for Astrophysics, Italy and CEA (Paris), France.
This work made use of THINGS, `The H$_\mathrm{I}$ Nearby Galaxy Survey’ \citep{Walter2008} and HERACLES, `The HERA CO-Line Extragalactic Survey’ \citep{Leroy2009}.
This research made use of \texttt{Astropy}: a community-developed core Python package and an ecosystem of tools and resources for astronomy \citep[][\url{http://www.astropy.org}]{astropy:2013, astropy:2018, astropy:2022}; \texttt{matplotlib}, a Python library for publication quality graphics \citep{Hunter2007}; \texttt{NumPy} \citep{Harris2020}; \texttt{SciPy} \citep{SciPy-NMeth}; TOPCAT, an
interactive graphical viewer and editor for tabular data
\citep{Taylor2005}; \texttt{Photutils}, an \texttt{Astropy} package for detection and photometry of astronomical sources \citep{Bradley2024}.

\end{acknowledgements}

\bibliographystyle{aa}
\bibliography{References}

\appendix
\onecolumn

\section{Maps of galaxies' properties} \label{ap:maps}

In this Appendix we show maps of the resolved properties of galaxies in our sample, at a resolution of 18".

   \begin{figure*}[!htbp]
   \centering
   \includegraphics[width=0.9\textwidth]{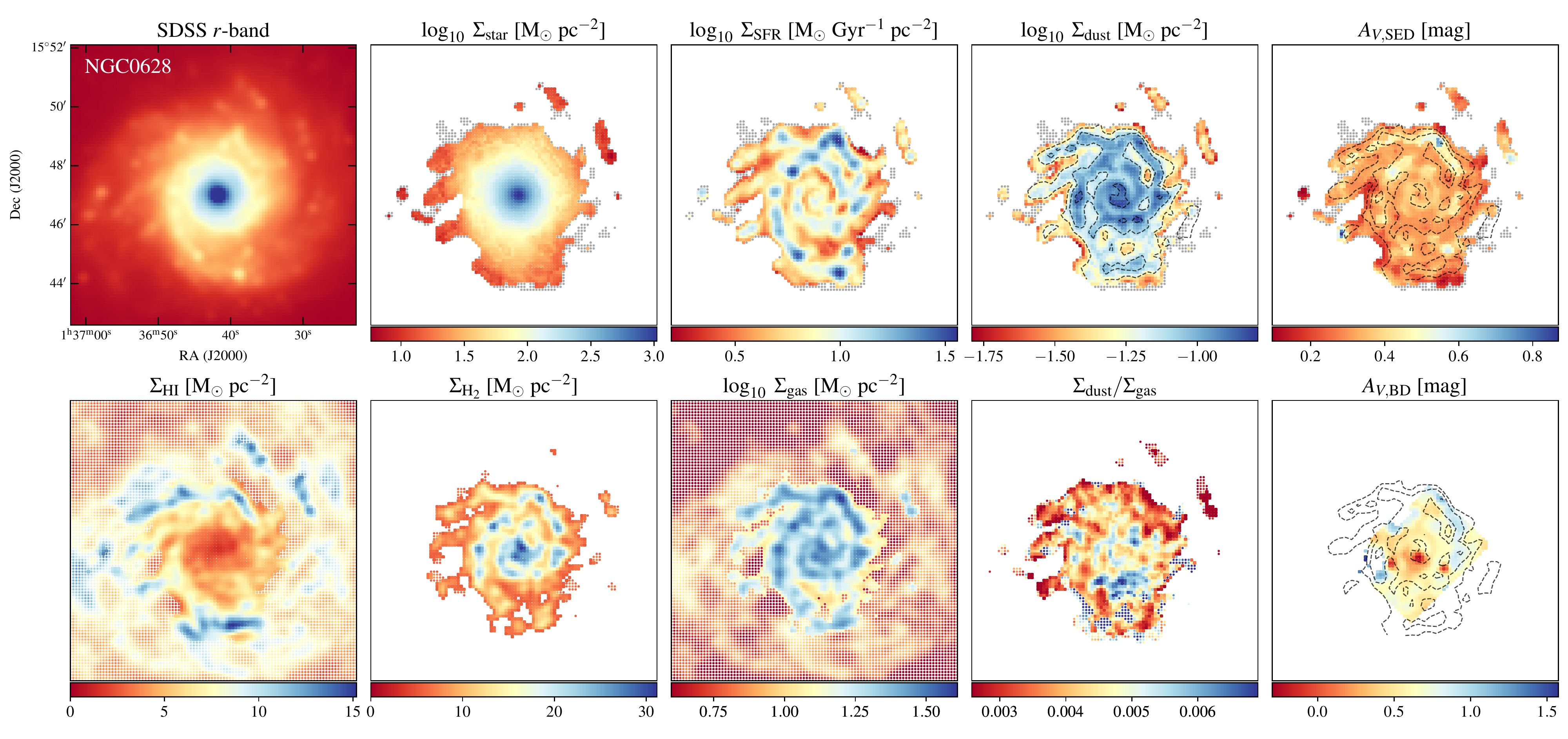}
   \caption{Maps of observed and derived properties of NGC~0628. \textit{Top}: Convolved and regridded SDSS r-band, $\log_{10}\Sigma_\mathrm{star}$, $\log_{10}\Sigma_\mathrm{SFR}$, $\log_{10}\Sigma_\mathrm{dust}$, as well as SED-fitting derived V-band attenuation ($A_{V,\mathrm{SED}}$), from left to right.
   \textit{Bottom}: $\Sigma_\mathrm{HI}$, $\Sigma_\mathrm{H2}$, $\log_{10}\Sigma_\mathrm{gas}$, DGR and V-band attenuation derived by the BD ($A_{V,\mathrm{BD}}$), from left to right. 
   Gray points correspond to pixels that are rejected (see Sec.~\ref{sec:steps}, ~\ref{sec:fitting} for more details). The $\Sigma_\mathrm{HI}$ and $\Sigma_\mathrm{H2}$ maps extend up to their corresponding 3$\sigmaup$ limit. 
   For $\Sigma_\mathrm{HI}$, $\Sigma_\mathrm{H2}$ and $A_{V,\mathrm{BD}}$ maps pixels that correspond to areas excluded by the SED-fitting analysis are plotted with smaller dots. 
   In $\log_{10}\Sigma_\mathrm{gas}$ map, pixels excluded by the SED-fitting analysis and having both \textsc{Hi} and CO detection are depicted by smaller squares, while pixels with only \textsc{Hi} and not CO are plotted with dots.   
   Contours plotted in some map are from $\log_{10}$~$\Sigma_\mathrm{dust}$ [M$_\odot$~pc$^{-2}$] maps with a lowest contour at -1.3 and linear spacing with the highest at -0.9. Dots in each map indicate areas excluded fro the SED fitting routine.}
   \label{fig:mapsNGC0628}
   \end{figure*}
 
   \begin{figure*}[!htbp]
   \centering
   \includegraphics[width=0.9
   \textwidth]{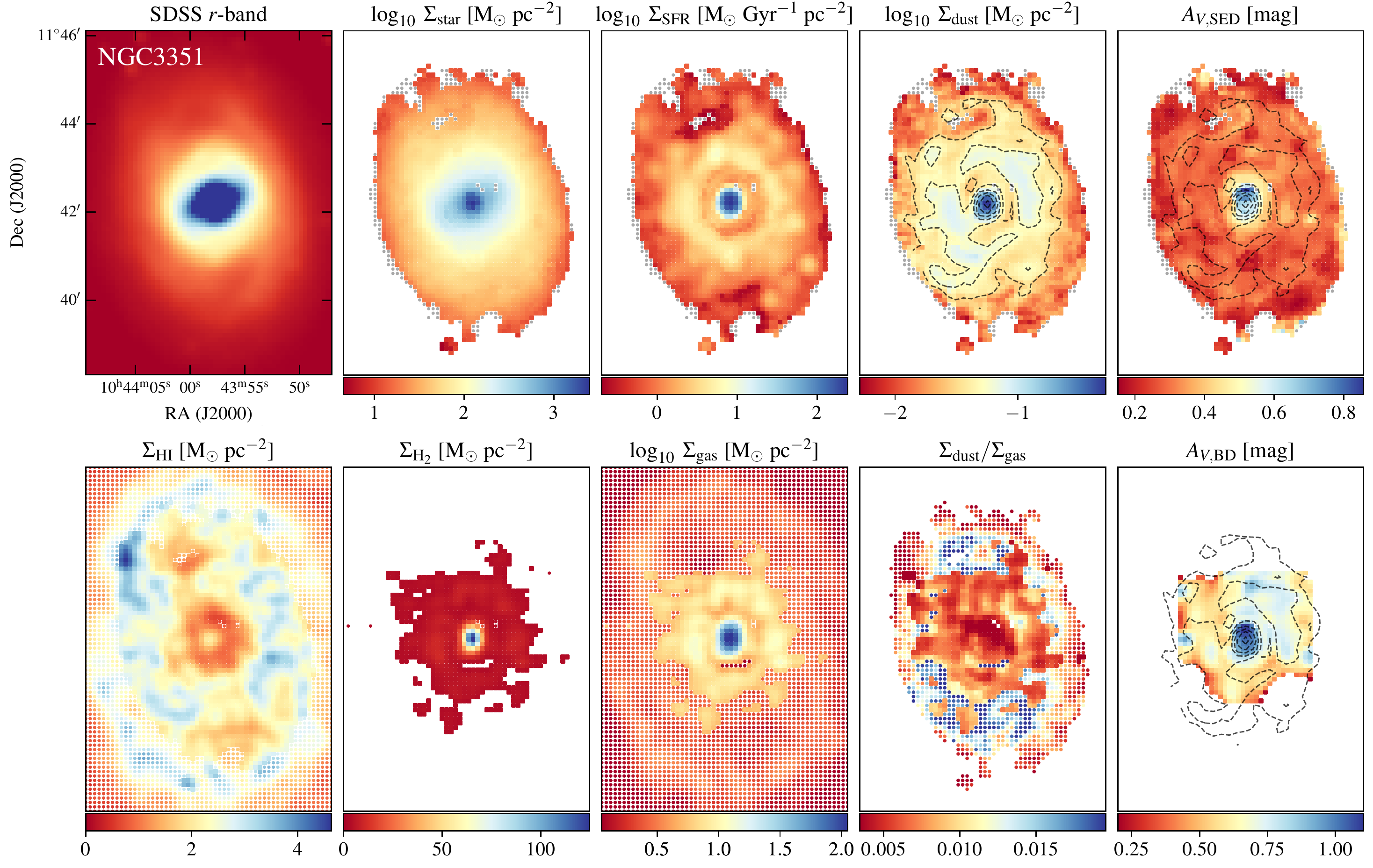}
   \caption{Same as Fig.~\ref{fig:mapsNGC0628}, but for NGC~3351. Contours are taken from the $\log_{10}$~$\Sigma_\mathrm{dust}$ [M$_\odot$~pc$^{-2}$] maps with a lowest contour at -1.5 and linear spacing with the highest at 0.}
   \label{fig:mapsNGC3351}
   \end{figure*}

   \begin{figure*}[!htbp]
   \vspace{1.0 cm}\centering
   \includegraphics[width=0.8\textwidth]{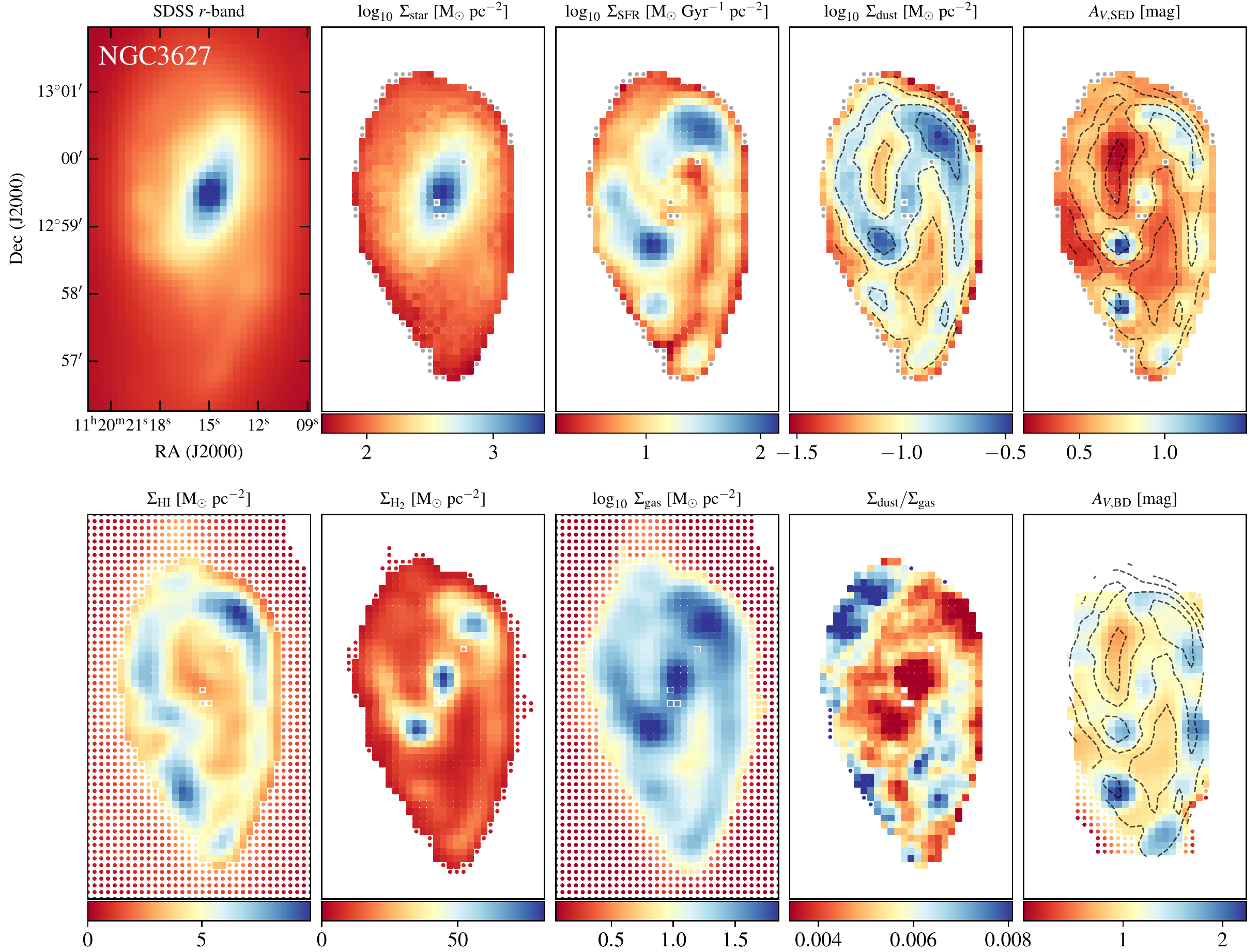}
   \caption{Same as Fig.~\ref{fig:mapsNGC0628}, but for NGC~3627. Contours are taken from the $\log_{10}$~$\Sigma_\mathrm{dust}$ [M$_\odot$~pc$^{-2}$] maps with a lowest contour at -1.5 and linear spacing with the highest at -0.7.}
   \label{fig:mapsNGC3627}
   \end{figure*}

   \begin{figure*}[!htbp]
   \vspace{1.8 cm}\centering
   \includegraphics[width=\textwidth]{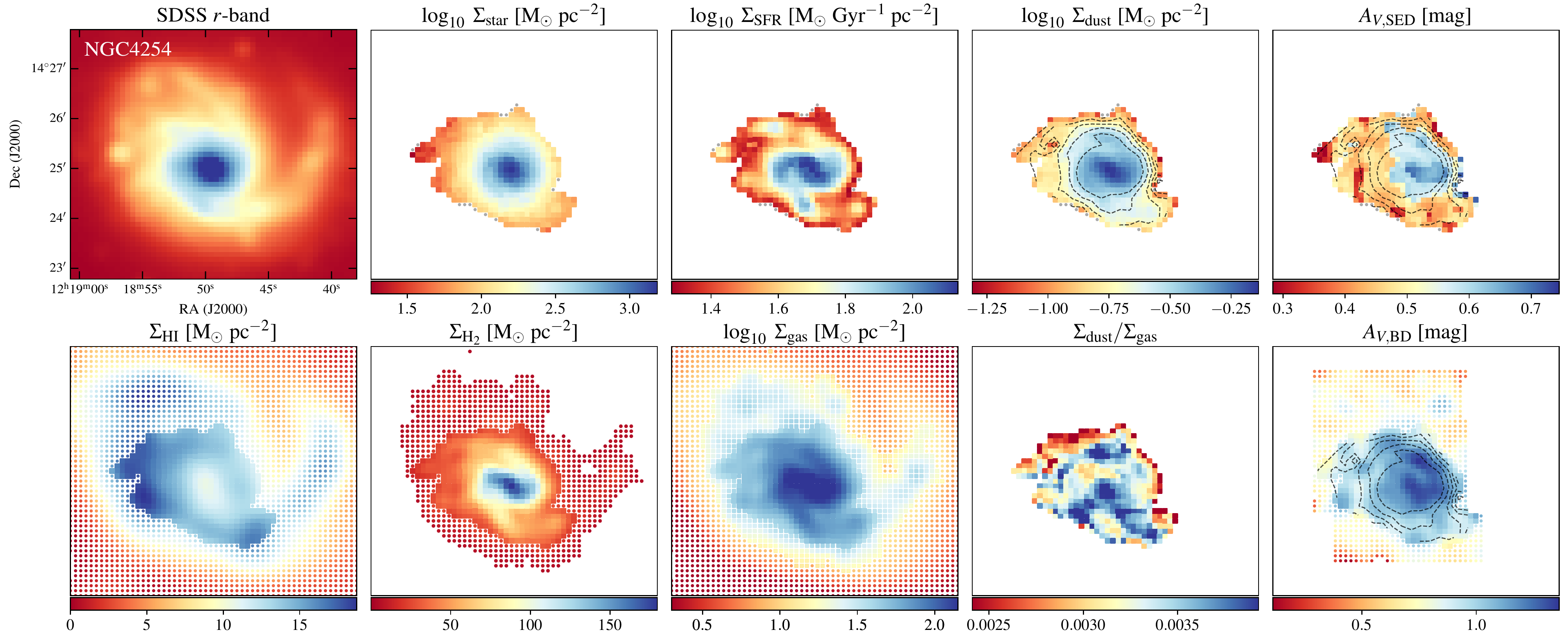}
   \caption{Same as Fig.~\ref{fig:mapsNGC0628}, but for NGC~4254. Contours are taken from the $\log_{10}$~$\Sigma_\mathrm{dust}$ [M$_\odot$~pc$^{-2}$] maps with a lowest contour at -1.1 and linear spacing with the highest at -0.5.}
   \label{fig:mapsNGC4254}
   \end{figure*}
   
   \begin{figure*}[t!]
   \centering
   \includegraphics[width=\textwidth]{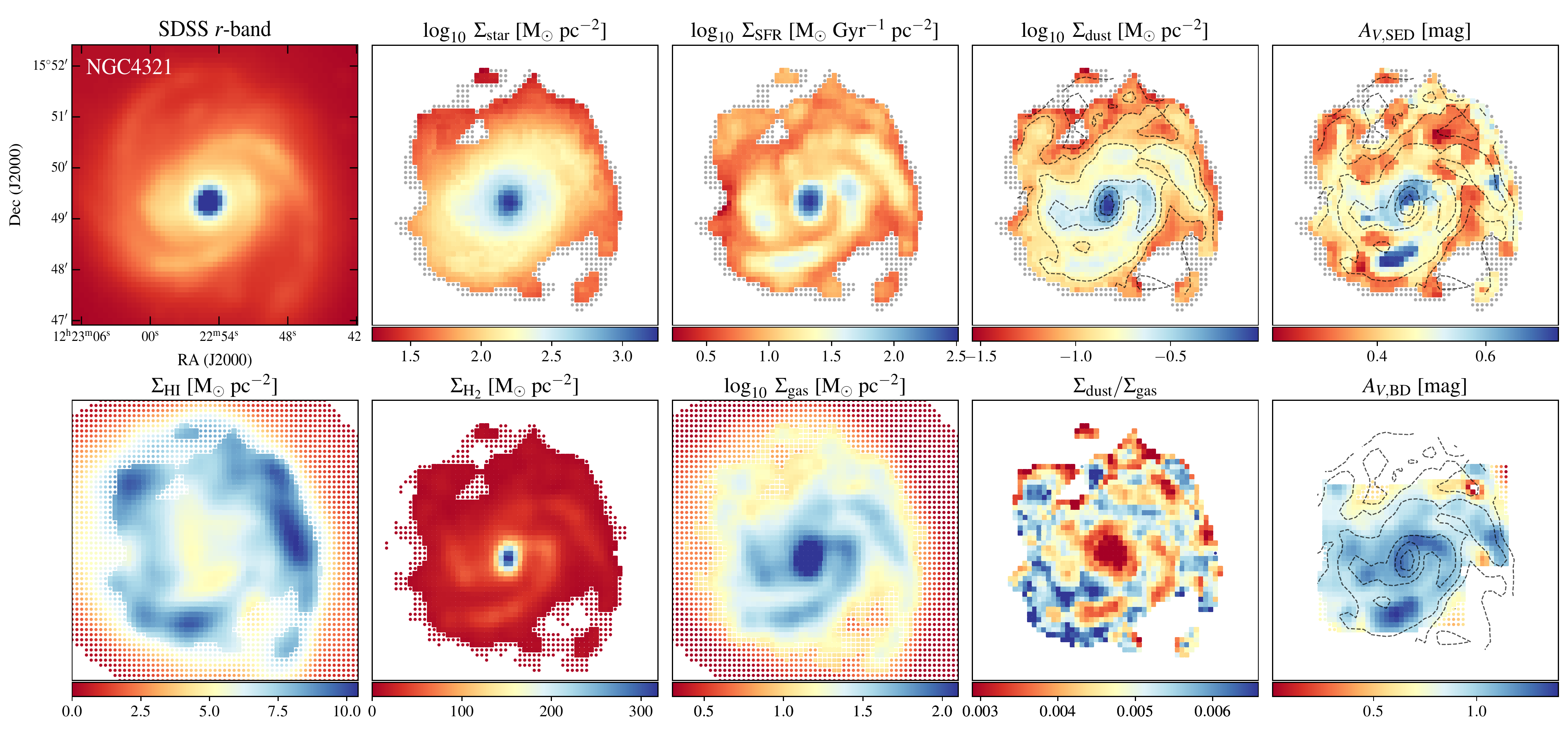}
   \caption{Same as Fig.~\ref{fig:mapsNGC0628}, but for NGC~4321.
   Contours are taken from the $\log_{10}$~$\Sigma_\mathrm{dust}$ [M$_\odot$~pc$^{-2}$] maps with a lowest contour at -1.5 and linear spacing with the highest at 0.
   }
   \label{fig:apNGC4321}
   \end{figure*}

\section{CO-to-H$_{2}$ conversion factor $\alphaup_\mathrm{CO}$ and CO(2-1)-to-CO(1-0) line ratio $R_{21}$}
\label{ap:alpha_co}

In this Appendix, we discuss our assumptions to infer the molecular gas surface density from $^{12}$CO J=2-1 line maps and possible dependencies of our results from these assumptions. As stated in the main paper we used a MW-disc CO-to-H$_{2}$ conversion factor a$_\mathrm{CO}$= 4.4~M$_{\odot}$~pc$^{-2}$~(K km s$^{-1}$)$^{-1}$ \citep{Bolatto2013} and an intrinsic line ratio $R_{21}$=$I_{\mathrm{CO}(2-1)}$/$I_{\mathrm{CO}(1-0)}$=0.7 \citep[e.g.,][]{Leroy2009, Schruba2011}. Although these values have been widely used by previous studies of the ISM in nearby galaxies \citep[e.g.,][]{Casasola2017}, variations of these two parameters are possible due to variations of local ISM conditions in clouds, such as metallicity, temperature, mass surface density, velocity dispersion, optical thickness, etc. The metallicity dependencies of $\alphaup_{\mathrm{CO}}$ have been studied by many authors \citep[e.g.,][]{Amorin2016} but for our sample metallicities are very close to solar \citep{Brazzini2024} and we don't expect significant deviations from the assumed value across the bright discs we investigate. Other studies have found $\alphaup_{\mathrm{CO}}$ to deviate from the typical MW  in galaxy's central regions \citep[e.g.,][]{Israel2020, Teng2022}.

Dedicated surveys for accurate determination of $\alphaup_{\mathrm{CO}}$ often include galaxies of our sample, such as the CO-isotopologues investigation by \citet{Cormier2018} or the dust based studies by \citet{Sandstrom2013} and \citet{Chiang2024}. The usage of CO isotopologues as tracers of molecular mass primarily traces dense gas, leading to underestimation of the total molecular mass, given the presence of diffuse molecular gas in the ISM. \citet{Sandstrom2013} used the dust as gas tracer and the gas-to-dust ratio to infer $\alphaup_{\mathrm{CO}}$ in kpc-scale regions. They found an average $\alphaup_{\mathrm{CO}}$ radial profile that is generally flat, with a depression towards the centre, particularly evident for NGC~3351, NGC~3627 and NGC~4321 ($R$~<~0.1$R_{25}$). In the same area, the $\alphaup_{\mathrm{CO}}$ was found significantly increased in NGC~4254. They provided average $I_\mathrm{CO}$-weighted values for each galaxy and as optimal average $\alphaup_{\mathrm{CO}}$ for the whole sample, 3.1~M$_{\odot}$~pc$^{-2}$~(K km s$^{-1}$)$^{-1}$.
\citet{Chiang2024} using a similar method, studies the variation of $\alphaup_{\mathrm{CO}}$ as a function of several physical quantities, and find negative anti-correlations with SFR, radiation field intensity, CO intensity and galactocentric distance.

Galaxies in our sample have also been targets for $R_{21}$ investigations. Although these are also discussed by \citet{Sandstrom2013}, we use the data of more recent and dedicated surveys which give resulting values more consistent between each other and with our assumed ratio \citep{denBrok2021, Yajima2021, Leroy2022}. These works find $R_{21}$ in the range 0.45 -- 0.83 as average values across the disc of galaxies in our sample. The mean of the three values (two only for NGC~3351) reported for each galaxy by \citet{denBrok2021}, \citet{Yajima2021} and by \citet{Leroy2022} are 0.56, 0.67, 0.50, 0.69, 0.72 for NGC~0628, NGC~3351, NGC~3627, NGC~4254, and NGC~4321, respectively. These are within 30$\%$ of our assumed $R_{21}$ ratio, with NGC~4254 and NGC~4321 fully consistent with our assumed $R_{21}$ ratio. There are no significant radial variations measured for the region of interest in galaxies of our sample. Only for NGC~3627 and NGC~4321 the ALMA observations analysed by \citet{denBrok2021} show a positive radial gradient or a factor 1.5 enhancement in the centre, respectively. These radial variations are not however confirmed by other observations such as HERACLES (IRAM-30m) data. We notice that for NGC~3627 and NGC~4321 results of different analyses give somewhat different mean values. Because of these discrepancies we have considered a uniform $R_{21}$~=~0.7 ratio in the present study.

\begin{figure*}[t]
   \centering
   \includegraphics[width=\textwidth]{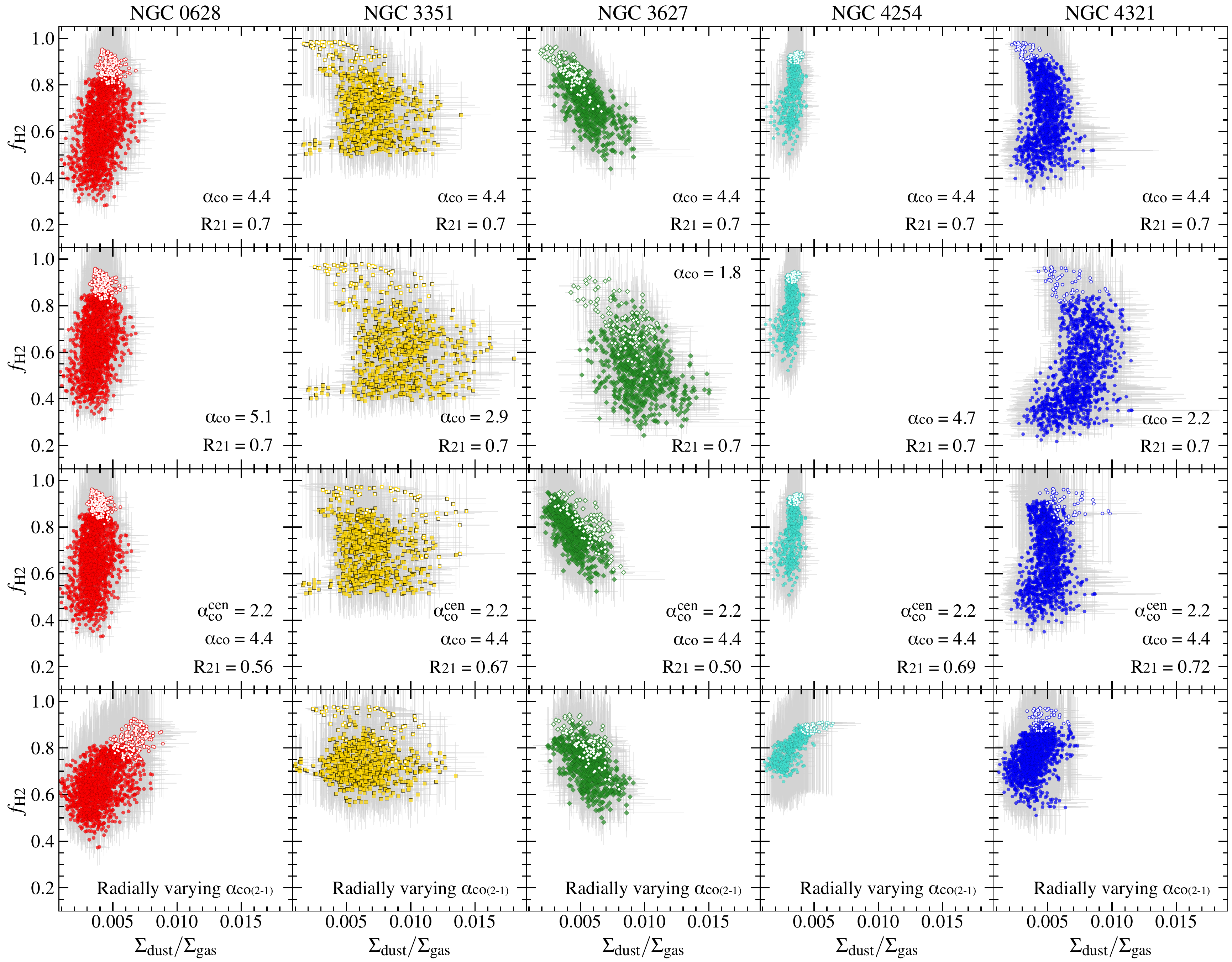}
   \caption{Molecular gas mass fraction as a function dust-to-gas mass ratio, for each galaxy in the sample. The molecular gas is estimated assuming a constant MW $\alphaup_\mathrm{CO}$~=~4.4~M$_\odot$~pc$^{-2}$~(K~km~s$^{-1}$)$^{-1}$ (\citealt{Bolatto2013} and $R_{21}$=0.7 (\textit{top panels}), 
   a radially constant $\alphaup_\mathrm{CO}$ which varies from galaxy to galaxy as found by \citet{Sandstrom2013} (\textit{middle-top panels}), a constant $\alphaup_\mathrm{CO}$~=~4.4~M$_\odot$~pc$^{-2}$~(K~km~s$^{-1}$)$^{-1}$ throughout the galaxies except in the central areas, and a radially constant $R_{21}$ value varying from galaxy to galaxy \citep[][\textit{middle-bottom panels}]{denBrok2021,Yajima2021,Leroy2022}, and a radially varying $\alphaup_\mathrm{CO(2-1)}$ as found by \citet[][\textit{bottom panels}]{Chiang2024}. The $\alphaup_\mathrm{CO}$ and $R_{21}$ values used for each panel are shown in the lower-right corners. Pixels within the inner 0.15~$R_{25}$ area are plotted with open symbols. Uncertainties are shown with gray error-bars.}
   \label{fig:alpha_co}
   \end{figure*}

With the aim to verify if our results are significantly affected by variations of $\alphaup_{\mathrm{CO}}$ and $R_{21}$ with respect to the assumed values, in Fig.~\ref{fig:alpha_co} we plot the $f_\mathrm{H2}$ as a function of the DGR for four different cases. 
In the top row, we assume the typical $\alphaup_{\mathrm{CO}}$~=~4.4~M$_{\odot}$~pc$^{-2}$~(K km s$^{-1}$)$^{-1}$ as in the rest of the paper (same as Fig.~\ref{fig:fH2} but with a common x-axis range for all galaxies); in the second row, the average $\alphaup_{\mathrm{CO}}$ values found by \citet{Sandstrom2013} for each galaxy with a constant $R_{21}$ has been used; in the third row, we set $\alphaup_{\mathrm{CO}}$ = 4.4~M$_{\odot}$~pc$^{-2}$~(K km s$^{-1}$)$^{-1}$ and use $R_{21}$~=0.56, 0.67, 0.50, 0.69, 0.72 for NGC~0628, NGC~3351, NGC~3627, NGC~4254, and NGC~4321, respectively. In addition, for NGC~3351, NGC~3627 and NGC~4321 we use a 1.5 enhancement factor in $R_{21}$ and a factor 2 decrease in $\alphaup_{\mathrm{CO}}$ for the central region; finally, in the bottom row, we use the correlation found by \citet{Chiang2024} between $\alphaup_{\mathrm{CO(2-1)}}$ and galactocentric distance (with no need to correct for $R_{21}$). The pixels that lie in the $R$~<~0.15$R_{25}$ areas are plotted with open symbols.

   \begin{figure*}[t]
   \centering
   \includegraphics[width=\textwidth]{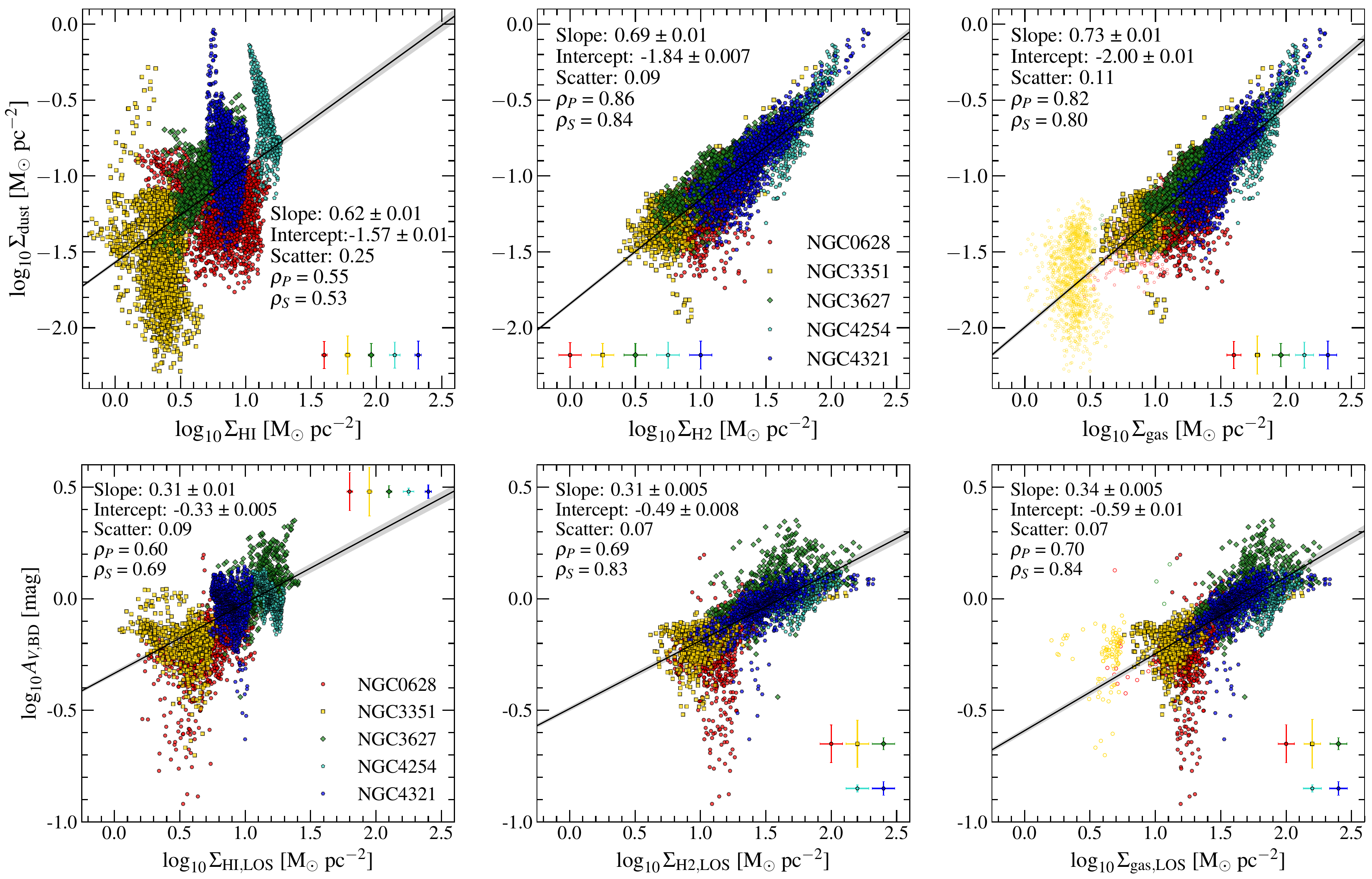}
   \caption{\textit{Top panels}, same as in Fig.~\ref{fig:dust-gas} and \textit{bottom panels} same as in Fig.~\ref{fig:Av-SigmaGas}, but with using the  \citet{Chiang2024} prescription for a radially varying $\alphaup_{\mathrm{CO}}$ (see text for details.)}
   \label{fig:alphaCOs}
   \end{figure*}

The main conclusion of our investigation in this Appendix is that by changing our assumptions on $R_{21}$ and on $\alphaup_{\mathrm{CO}}$ we recover similar DGR trends across the galaxies and from one galaxy to another. Although molecular fractions change, these always increase with the DGR for NGC~0628 and NGC~4254. If we exclude the central area, for which the CO-to-H$_2$ conversion factor can be lower than what has been shown in Fig.~\ref{fig:alpha_co}, NGC~3351 and NGC~4321 have no correlation between the molecular fraction and the DGR where the molecular phase dominates, while NGC~3627 shows a decreasing DGR as the molecular fraction increases. NGC~4321 shows a decrease of dust abundance in the outskirts, under all assumptions apart from the case of the radially-depended $\alphaup_{\mathrm{CO}}$ by \citet{Chiang2024}. In this case the main part of the disc shows a mild positive correlation, with some deviation for outskirts and centre. In an increasing order we list the average DGRs found in the sampled regions: 0.0033~$\pm$~0.0005, 0.0041~$\pm$~0.001, 0.0048~$\pm$~0.0010, 0.0054~$\pm$~0.0013 and 0.0069~$\pm$~0.0020, for NGC~4254, NGC~0628, NGC~4321, NGC~3627 and NGC~3351, respectively, using the standard values as in the main paper. These values increase by decreasing $\alphaup_{\mathrm{CO}}$ but the sequence is unchanged; also when using the \citealt{Chiang2024} $\alphaup_{\mathrm{CO}}$ prescription, with the only exception of NGC~4321 which in this case has a DGR equal to 0.0039~$\pm$~0.001 and is second in  increasing order, instead of being third  (see Sec.~\ref{sec:shield}).

We have also estimated the best-fit parameters and the correlation coefficients for the $\log_{10}\Sigma_\mathrm{dust}$--$\log_{10}\Sigma_\mathrm{H2}$ and $\log_{10}\Sigma_\mathrm{dust}$--$\log_{10}\Sigma_\mathrm{gas}$ relations for various $\alphaup_{\mathrm{CO}}$ and $R_{21}$ and we do not find significant differences. For instance, if we use the \citet{Chiang2024} prescription where $\alphaup_{\mathrm{CO}}$ changes with galactocentric distance, the correlations of $\log_{10}\Sigma_\mathrm{dust}$ with $\log_{10}\Sigma_\mathrm{H2}$ and $\log_{10}\Sigma_\mathrm{gas}$, as well as $\log_{10}A_{V,\mathrm{BD}}$ with $\log_{10}\Sigma_\mathrm{H2}$ and $\log_{10}\Sigma_\mathrm{gas}$, have similar slopes, scatter and correlation significance to those found using a constant $\alphaup_{\mathrm{CO}}$ (see Fig.~\ref{fig:alphaCOs}). We prefer to use the constant $\alphaup_{\mathrm{CO}}$ values in the main paper to avoid any possible influence of circular reasoning: in fact the \citet{Chiang2024} prescription is found assuming that the total gas surface density is traced by the dust surface density given a constant gas-to-metal ratio. Thus we would study the correlation of $\Sigma_\mathrm{dust}$ with a quantity that has been derived using $\Sigma_\mathrm{dust}$ (though assuming a different dust model than ours, as in \citealt{Chiang2024}). Nevertheless, similar correlations (not shown here) are found also using the \citet{Bolatto2013} formula that accounts for a decrease of $\alphaup_{\mathrm{CO}}$ in denser regions (see their Eq. 31). This is recovered using several methodologies and not only using dust as a proxy of gas. In any case, the $\alphaup_{\mathrm{CO}}$ variations affects mostly central areas, that are only a small portion of the discs  we investigate, and do not affect our main conclusions.

\section{Attenuations for simple geometries}
\label{ap:atte}

\begin{figure*}[!htbp]
\centering
\includegraphics[width=0.5\textwidth]{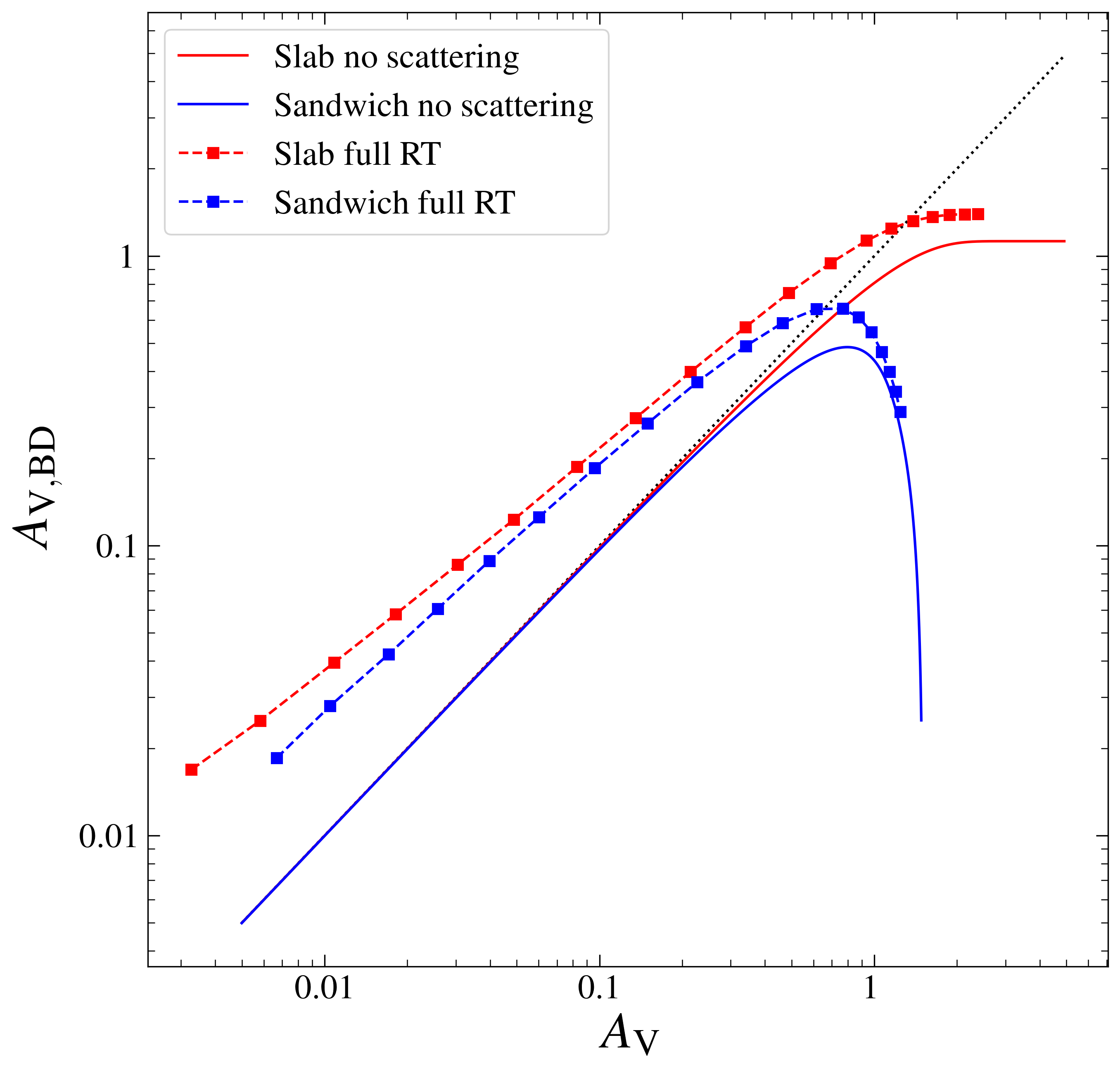}
\caption{$A_\mathrm{V, BD}$ vs $A_V$ for the models discussed in the text. The dotted line shows the 1-to-1 relation.}
\label{fig:atte}
\end{figure*}

For a homogeneously mixed thin layer of dust and sources, of optical depth $\tau_\lambda$ perpendicular to the layer, the radiative transfer solution can be derived analytically in the absence of scattering: it is the {\em slab} model \citep{Disney1989} and the corresponding attenuation 
can be written as  
\begin{equation}
A_\lambda=-2.5 \log_{10} \frac{I}{I_0} = -2.5 \log_{10} \frac{1-e^{-\tau_\lambda}}{\tau_\lambda},
\label{eq:slab}
\end{equation}
where, for simplicity, we consider only the line of sight perpendicular to the layer (face-on view). 
For $\tau_\lambda \rightarrow 0$, Eq.~\ref{eq:slab} can be written using series expansion as
\begin{equation}
A_\lambda=1.086\frac{\tau_\lambda}{2}.
\end{equation}
Instead, for $\tau\rightarrow\infty$, it is
\begin{equation}
A_\lambda=2.5\log_{10} \tau_\lambda.
\end{equation}
Using these limits, $\tau_\lambda = k(\lambda) \tau_\mathrm{V}$ with $k(\lambda)$ the (V-band normalised) extinction law, and Eq.~\ref{eq:avbd}, rewritten as
\begin{equation}
A_\mathrm{V, BD} = \frac{A_{\mathrm{H}_\betaup} - A_{\mathrm{H}_\alphaup}
}{k(\mathrm{H}\betaup) - k(\mathrm{H}\alphaup)},
\end{equation}
 we can easily find that 
$A_\mathrm{V, BD}$ tends to $A_\mathrm{V}$ for low optical depths, and to the constant value $2.5\log_{10}(k(\mathrm{H}\betaup)/k(\mathrm{H}\alphaup))/(k(\mathrm{H}\betaup)-k(\mathrm{H}\alphaup))$ at high optical depths. This is shown in Fig.~\ref{fig:atte}, where we plot $A_\mathrm{V,BD}$ vs $A_\mathrm{V}$ for a \textit{slab} 
using the extinction law of the THEMIS dust model ($k(\mathrm{H}\alphaup)=0.81$, $k(\mathrm{H}\betaup)=1.14$).

For the {\em sandwich} model, it is 
\begin{equation}
A_\lambda=-2.5 \log_{10} \frac{I}{I_0} = -2.5 \log_{10} \left(
\frac{h_d}{h_s}\frac{1-e^{-\tau_\lambda}}{\tau_\lambda}
+\frac{1}{2}(1-\frac{h_d}{h_s})e^{-\tau_\lambda}
+\frac{1}{2}(1-\frac{h_d}{h_s})
\right),
\label{eq:sandwich}
\end{equation}
where $h_s$ is the thickness of the source layer and $h_d \le h_s$ is that of a dust layer internal to the source one and sharing the same mid-plane:
the solution is the combination of a \textit{slab} model, of the dust-free portion of sources on the side opposite to the viewer attenuated by the dust screen, and of the unattenuated symmetric portion on the side of the viewer (the first, second and third terms of the logarithm in Eq.~\ref{eq:sandwich}, respectively; the \textit{sandwich} is equivalent to the \textit{slab} for $h_d = h_s$). As for the \textit{slab}, $A_\mathrm{V, BD}$ is the same as $A_\mathrm{V}$ in the optically thin limit. Instead, it goes to zero as $1/\tau_V$ at high $A_\mathrm{V}$. The case for $h_d=h_s/2$ is shown in Fig.~\ref{fig:atte}.

For the \textit{slab} and \textit{sandwich} models we also made full radiative transfer calculation including scattering:
we used the TRADING code \citep{Bianchi2008}, assuming $h_d=h_s/2$ and albedos from the THEMIS model [$\omega(\mathrm{H}\alphaup)=0.56$, $\omega(\mathrm{H}\betaup)=0.52$]; and strong forward scattering, with an asymmetry parameter characteristic for MW dust, $g=0.7$,
at both wavelengths \citep{Hensley2021}. We run 19 models with face-on optical depth $\tau_V$ between 0.1 and 20. 

Results from radiative transfer calculations are also shown in Fig.~\ref{fig:atte}. The trends are qualitatively similar to the no-scattering cases, but with differences between $A_\mathrm{V, BD}$ and $A_\mathrm{V}$ also in the optically thin limit. The full radiative transfer results for the \textit{slab} and \textit{sandwich} models are also discussed in Sect.~\ref{sec:avsd} and plotted in Fig.~\ref{fig:Av-Dust}.

\end{document}